\documentclass[twocolumn,english,superscriptaddress,showpacs,aps,prb]{revtex4}
\usepackage[T1]{fontenc}
\usepackage[latin9]{inputenc}
\usepackage{bm}
\usepackage{amsmath}
\usepackage{amssymb}
\usepackage{graphicx}
\usepackage{esint}

\makeatletter
\@ifundefined{textcolor}{}
{%
 \definecolor{BLACK}{gray}{0}
 \definecolor{WHITE}{gray}{1}
 \definecolor{RED}{rgb}{1,0,0}
 \definecolor{GREEN}{rgb}{0,1,0}
 \definecolor{BLUE}{rgb}{0,0,1}
 \definecolor{CYAN}{cmyk}{1,0,0,0}
 \definecolor{MAGENTA}{cmyk}{0,1,0,0}
 \definecolor{YELLOW}{cmyk}{0,0,1,0}
}

\usepackage{hyperref}
\hypersetup{
breaklinks=true,
colorlinks=true,
citecolor=blue,
linkcolor=blue,
filecolor=blue,
urlcolor=blue
}

\usepackage{psfrag}
\usepackage[usenames,dvipsnames]{color}

\makeatother

\usepackage{babel}
\begin{document}

\title{Dissipation effects in random transverse-field Ising chains}

\author{José A. Hoyos}

\affiliation{Instituto de F\'{i}sica de São Carlos, Universidade de São Paulo,
C.P. 369, São Carlos, São Paulo 13560-970, Brazil}

\author{Thomas Vojta}

\affiliation{Department of Physics, Missouri University of Science and Technology,
Rolla, MO 65409, USA}
\begin{abstract}
We study the effects of Ohmic, super-Ohmic, and sub-Ohmic dissipation
on the zero-temperature quantum phase transition in the random transverse-field
Ising chain by means of an (asymptotically exact) analytical strong-disorder
renormalization-group approach. We find that Ohmic damping destabilizes
the infinite-randomness critical point and the associated quantum
Griffiths singularities of the dissipationless system. The quantum
dynamics of large magnetic clusters freezes completely which destroys
the sharp phase transition by smearing. The effects of sub-Ohmic dissipation
are similar and also lead to a smeared transition. In contrast, super-Ohmic
damping is an irrelevant perturbation; the critical behavior is thus
identical to that of the dissipationless system. We discuss the resulting
phase diagrams, the behavior of various observables, and the implications
to higher dimensions and experiments. 
\end{abstract}

\pacs{05.10.Cc, 05.70.Fh, 75.10.-b}

\maketitle

\section{Introduction\label{sec:Introduction} }

Continuous phase transitions display the remarkable feature of universality:
The physics sufficiently close to the transition point is independent
of microscopic details, it only depends on a small number of key parameters
such as the symmetry of the order parameter, the dimensionality of
the system, and the presence or absence of frustration (for reviews
see, e.g., Refs.\ \onlinecite{Goldenfeld-book92,fisher-rmp98}).
For this reason, simple prototypical models are extensively used in
theoretical studies of continuous phase transitions as their critical
behavior can be expected to exactly reproduce that of experimental
systems.

Because realistic systems often contain considerable amounts of quenched
(time-independent) disorder (randomness), it is important to establish
whether or not such disorder is relevant or irrelevant at a continuous
phase transition. In other words, is disorder one of the unimportant
microscopic details or is it one of the key parameters that determine
the critical behavior? Interestingly, the answer to this question
is not unique but depends on the transition at hand. Early insight
was gained from analyzing the fate of the disorder strength under
coarse graining: According to the Harris criterion,\ \cite{harris-jpc74}
disorder is perturbatively relevant at a critical point if its clean
correlation length exponent $\nu$ fulfills the inequality $d\nu<2$
with $d$ the space dimensionality. In this case, the critical behavior
of the disordered system must differ from that of the clean one. However,
the ultimate fate of the transition can not be inferred from the Harris
criterion.

In recent years, it has become increasingly clear that rare, strongly
interacting spatial regions play an important role in phase transitions
in disordered systems. These regions can be (locally) in the ordered
phase while the bulk system is still in the disordered phase. Their
dynamics becomes very slow because it involves coherently changing
the order parameter in a large volume. Griffiths\ \cite{griffiths-prl69,mccoy-prl69}
showed that these rare regions lead to non-analyticities in the free
energy not just at the critical point but in an entire parameter region
around the transition which is now known as the Griffiths phase. At
generic thermal (classical) transitions, the resulting Griffiths singularities
in thermodynamic quantities are very weak and probably unobservable
in experiment.\ \cite{imry-prb77}

Rare regions are more important for zero-temperature quantum phase
transitions, the critical behavior of which is determined by order-parameter
fluctuations in space and time. As quenched disorder is perfectly
correlated in the time direction, disorder effects are enhanced. The
resulting strong quantum Griffiths singularities\ \cite{thill-huse-physa95,guo-bhatt-huse-prb96,young-rieger-prb96}
of thermodynamic quantities can take power-law form. The critical
point itself remains sharp but can be of exotic infinite-randomness
type as was shown by Fisher.\ \cite{fisher92,fisher95}

At quantum phase transitions, statics and dynamics are intimately
coupled. Therefore, changes in the quantum dynamics can cause changes
in the thermodynamic behavior. For example, dissipation, i.e., damping
of the order-parameter fluctuations due to additional degrees of freedom,
can further enhance the rare-region effects. In systems with Ising
symmetry, each rare region acts as a two-level system. In the presence
of Ohmic dissipation, it undergoes the localization quantum phase
transition of the spin-boson problem.\ \cite{leggett-etal-rmp87}
Thus, the quantum dynamics of sufficiently large rare regions freezes,\ \cite{millis-morr-schmalian-prl01,millis-morr-schmalian-prb02}
leading to a smearing of the global phase transition.\ \cite{vojta-prl03}

In order to gain a more complete understanding of the non-perturbative
physics of these dissipative rare regions, the heuristic arguments
of Refs.\  \onlinecite{millis-morr-schmalian-prl01,millis-morr-schmalian-prb02,vojta-prl03}
need to be complemented by an explicit calculation of a prototypical
microscopic model. Schehr and Rieger made significant progress in
this direction by applying a numerical strong-disorder renormalization
group to the dissipative transverse-field Ising model.\ \cite{schehr-rieger06,scherh-rieger-jsm08}
Their computer simulation results supported the above smeared-transition
scenario but focused on the infinite-randomness physics arising at
intermediate energies.

In this paper, we derive a comprehensive analytical approach to the
quantum phase transition of the transverse-field Ising chain coupled
to dissipative baths of harmonic oscillators. The method is a generalization
of Fisher's solution\ \cite{fisher92,fisher95} of the dissipationless
case which treats on equal footing the effects of dissipation and
disorder. For Ohmic damping, the theory captures the full crossover
between the infinite-randomness (and quantum Griffiths) physics at
higher energies where the dissipation is less important and the dissipation-dominated
smeared transition at low energies. The effects of sub-Ohmic dissipation
are qualitatively similar and also result in a smeared transition.
In contrast, super-Ohmic dissipation is irrelevant, and the transition
is identical to that of the dissipationless chain. A short account
of part of these results was already published in Ref.\  \onlinecite{hoyos-vojta-prl08}.

Our paper is organized as follows: We define the model in Sec.\ \ref{sec:The-Hamiltonian}.
In Secs.\ \ref{sec:RG-recursion-relation} and \ref{sec:Flow-equations},
we derive the renormalization-group recursion relations and the corresponding
flow equations. They are solved in Sec.\ \ref{sec:Formal-solution}.
Section\ \ref{sec:Phase-diagram} is devoted to phase diagrams, observables,
and crossover effects. In Sec.\ \ref{sec:Discussions}, we discuss
the applicability of our method for weak disorder, generalizations
to higher dimensions, and the relevance of our results for experiments.
We conclude and compare with related results in the literature in
Sec.\ \ref{sec:Conclusions}.

\section{The Hamiltonian \label{sec:The-Hamiltonian}}

Our model Hamiltonian consists of a one-dimensional random transverse-field
Ising model coupled to a bosonic bath,\ \cite{schehr-rieger06} 
\begin{equation}
H=H_{I}+H_{B}+H_{C}.\label{eq:H}
\end{equation}
 Here, 
\begin{equation}
H_{I}=-\sum_{i}J_{i}\sigma_{i}^{z}\sigma_{i+1}^{z}-\sum_{i}h_{i}\sigma_{i}^{x}\label{eq:H-Ising}
\end{equation}
 is the Hamiltonian of the (dissipationless) random transverse-field
Ising chain. The spin-1/2 degree of freedom at site $i$ is described
by the Pauli matrices $\bm{\sigma}_{i}$, the local ferromagnetic
interaction is represented by $J_{i}$, and the local transverse field
$h_{i}$ controls the strength of the quantum fluctuations.

Dissipation, i.e., damping of the quantum fluctuations, is introduced
by coupling the $z$-component of each spin to an independent bath
of quantum harmonic oscillators.\ \cite{caldeira-leggett-prl81}
The bath Hamiltonian reads as 
\begin{equation}
H_{B}=\sum_{k,i}\omega_{k,i}\left(a_{k,i}^{\dagger}a_{k,i}^{\phantom{]\dagger}}+\frac{1}{2}\right),\label{eq:H-oscillators}
\end{equation}
 with $\omega_{k,i}$ being the frequency of the $k$-th harmonic
oscillator coupled to the spin at site $i$, and $a_{k,i}^{\phantom{]\dagger}}$
($a_{k,i}^{\dagger}$) being the usual annihilation (creation) operators.
(Note that we set $\hbar=1$.) Finally, the coupling between the spins
and their respective baths is given by 
\begin{equation}
H_{C}=\sum_{i}\sigma_{i}^{z}\sum_{k}\lambda_{k,i}\left(a_{k,i}^{\dagger}+a_{k,i}^{\phantom{\dagger}}\right),\label{eq:H-coupling}
\end{equation}
 with $\lambda_{k,i}$ being the strength of the interaction.

All relevant information about the bosonic baths is encoded in their
spectral densities ${\cal E}_{i}$, which can be parameterized as
\begin{equation}
{\cal E}_{i}(\omega)=\pi\sum_{k}\lambda_{k,i}^{2}\delta\left(\omega-\omega_{k,i}\right)=\frac{\pi}{2}\alpha_{i}\omega_{0,i}^{1-s}\omega^{s}\quad(\omega<\omega_{c,i}).\label{eq:spectral-function}
\end{equation}
 Here, $\omega_{c,i}$ is an ultraviolet cutoff energy, and $\alpha_{i}$
is a dimensionless measure of the dissipation strength. The microscopic
energy scale $\omega_{0,i}$ is often identified with the cutoff $\omega_{c,i}$.
However, we need to keep the two energies separate because the cutoff
$\omega_{c,i}$ will flow within our renormalization-group procedure,
while $\omega_{0,i}$ (the bare cutoff energy) will not. Depending
on the value of the exponent $s$, different types of dissipation
can be distinguished. $s=1$ corresponds to the experimentally important
Ohmic case. For $s>1$, the bath is super-Ohmic, which leads to qualitatively
weaker dissipation effects as there are fewer bath states at low energies.
For $s<1$, the bath is dubbed sub-Ohmic.

We emphasize that in the model (\ref{eq:H}), each spin is coupled
to its own (local) oscillator bath. Therefore, these baths damp the
local order-parameter fluctuations, i.e., they induce a long-range
interaction in time. However, they do not produce interactions between
different sites. If we coupled all spins to the same (global) oscillator
bath, we would also obtain an effective interaction, mediated by the
bath oscillators, between the spins at different sites. This interaction
would be analogous to the Ruderman-Kittel-Kasuya-Yosida (RKKY) interaction
between localized magnetic moments in metals.

Quenched disorder is implemented by taking all parameters in the Hamiltonian
(\ref{eq:H}), i.e., the interactions $J_{i}$, the transverse fields
$h_{i}$, the dissipation strengths $\alpha_{i}$, as well as the
bath energy scales $\omega_{c,i}$ and $\omega_{0,i}$ to be independent
random variables. Actually, making one of these quantities random
is sufficient because, under the renormalization-group flow (discussed
in Secs.\ \ref{sec:RG-recursion-relation} and \ref{sec:Flow-equations}),
the other quantities acquire randomness even if their bare values
are not random. In the following, we thus assume that in the bare
system only $J_{i}$ and $h_{i}$ are random, with probability distributions
$P_{I}(J)$ and $R_{I}(h)$, while $\alpha_{i}=\alpha_{I}\equiv\alpha$,
$\omega_{0,i}=\omega_{I}$, and $\omega_{c,i}=\omega_{c}$ are uniform.
We also restrict ourselves to the experimentally most interesting
case\ \cite{leggett-etal-rmp87} of the bath energy $\omega_{I}$
being the largest energy scale in the problem, $\omega_{I}\gg h_{i,}J_{i}$.

\section{Renormalization group recursion relations \label{sec:RG-recursion-relation}}

We now turn to the derivation of our theory. Our intent is to apply
a real-space-based strong-disorder renormalization-group method to
the Hamiltonian (\ref{eq:H}). This technique was introduced to tackle
random antiferromagnetic spin-1/2 chains\ \cite{MDH-PRL,MDH-PRB}
and has been successfully generalized to many disordered systems (for
a review, see Ref.\ \onlinecite{igloi-review}). Its philosophy is
to successively integrate out the degrees of freedom with the highest
local energies. This approach is justified \emph{a posteriori} if
the renormalized disorder strength becomes very large.

\subsection{Review of the dissipationless case}

The strong-disorder renormalization group of the dissipationless random
transverse-field Ising chain (\ref{eq:H-Ising}) was derived and solved
by Fisher.\ \cite{fisher92,fisher95} One starts by identifying the
highest local energy scale, $\Omega=\max\{h_{i},J_{i}\}$, in the
system.

If the largest energy is an interaction, say, $J_{2}$, the low-energy
states are those in which $\sigma_{2}^{z}$ and $\sigma_{3}^{z}$
are parallel: $\left|++\right\rangle $ and $\left|--\right\rangle $.
Therefore, the cluster of sites 2 and 3 can be recast as a single
effective spin-1/2 degree of freedom, the effective magnetic moment
of which is $\tilde{\mu}=\mu_{2}+\mu_{3}$. The transverse fields
$h_{2}$ and $h_{3}$ will lift the degeneracy between $\left|++\right\rangle $
and $\left|--\right\rangle $ in second order of perturbation theory,
yielding an effective transverse field $\tilde{h}=h_{2}h_{3}/J_{2}$
for the new spin.

In contrast, when the largest energy is a transverse field, say $h_{2}$,
the spin at site 2 is rapidly fluctuating between the $\left|+\right\rangle $
and $\left|-\right\rangle $ eigenstates (in $z$ direction). It hence
does not contribute to magnetization and can be decimated from the
chain. The neighboring spins $\sigma_{1}$ and $\sigma_{3}$ become
connected via an effective coupling $\tilde{J}=J_{1}J_{2}/h_{2}$,
also obtained in second order of perturbation theory.

This summarizes the renormalization-group recursion relations for
the dissipationless case. Note that the recursion relations are symmetric
under exchanging $h$ and $J$, reflecting the self-duality of the
Hamiltonian (\ref{eq:H-Ising}).

\subsection{Generalization to the dissipative case \label{subsec:SDRG-diss}}

The generalization of the strong-disorder renormalization group to
the full Hamiltonian (\ref{eq:H}) is not unique because the bosonic
baths can be incorporated in different ways.

One possibility is implemented in the numerical work of Refs.\ \onlinecite{schehr-rieger06,scherh-rieger-jsm08}.
Here, the renormalization-group step still consists of identifying
the largest of all the interactions and transverse fields, $\Omega=\max\{J_{i},h_{i}\}$.
Before decimating the corresponding high-energy degree of freedom,
one integrates out the high-energy bath oscillators of the involved
sites. This modifies the transverse fields and thus changes the recursion
relations. Furthermore, the bath cutoff $\omega_{c}$ becomes site
dependent. While this scheme appears to work very well numerically,
we found it hard to implement analytically as it yields nonanalytic
flow equations for the distributions of $J$, $h$, and $\mu$.

This difficulty is overcome by integrating out the oscillator degrees
of freedom together with the lattice modes (related to strong interactions
and transverse fields). We thus define the renormalization-group energy
scale as $\Omega=\max\{J_{i},h_{i},\omega_{c}/p\}$, where $p\gg1$
is a dimensionless constant, the importance of which will become clear
below in the context of adiabatic renormalization. A renormalization-group
step now consists of decimating either an interaction $J_{i}=\Omega$,
a transverse field $h_{i}=\Omega$, or an oscillator with frequency
$\omega_{k,i}=p\Omega$. Upon lowering the renormalization-group energy
scale from $\Omega$ to $\Omega-{\rm d}\Omega$, the bath cutoff $\omega_{c}$
is thus reduced by $p\,{\rm d}\Omega$ at \emph{all} sites. In this
sense, the renormalization of the baths is \emph{global}, and not
local as in the scheme of Refs.\ \onlinecite{schehr-rieger06,scherh-rieger-jsm08}.
This allows us to treat $\omega_{c}$ as a non-random variable.

As it turns out, this modified scheme is analytically tractable. In
the following subsections, we derive the resulting recursion relations.

\subsection{Adiabatic renormalization of the bosonic baths \label{subsec:adiabatic}}

In order to understand the effects of dissipation on the quantum fluctuations,
we first consider a single spin coupled to a bosonic bath, i.e., the
well-known spin-boson problem\ \cite{leggett-etal-rmp87} given by
\begin{equation}
H_{{\rm sb}}=-h\sigma^{x}+\sigma^{z}\sum_{k}\lambda_{k}\left(a_{k}^{\dagger}+a_{k}^{\phantom{\dagger}}\right)+\sum_{k}\omega_{k}\left(a_{k}^{\dagger}a_{k}^{\phantom{]\dagger}}+\frac{1}{2}\right).\label{eq:H-spin-boson}
\end{equation}
 The idea of the adiabatic renormalization is to eliminate high-frequency
oscillators in a Born-Oppenheimer type of approximation. Let us assume
that all oscillators of frequencies greater than $ph$ with $p\gg1$
can instantaneously follow the spin. We now wish to integrate out
these fast oscillators. To this end, consider the Hamiltonian 
\begin{align}
H_{0} & =\sigma^{z}\sum_{\omega_{k}>ph}\lambda_{k}\left(a_{k}^{\dagger}+a_{k}^{\phantom{\dagger}}\right)+\sum_{\omega_{k}>ph}\omega_{k}\left(a_{k}^{\dagger}a_{k}^{\phantom{]\dagger}}+\frac{1}{2}\right)\nonumber \\
 & =\sum_{\omega_{k}>ph}\omega_{k}\left(b_{k}^{\dagger}b_{k}^{\phantom{]\dagger}}+\frac{1}{2}\right),
\end{align}
 with $b_{k}^{\phantom{]\dagger}}=a_{k}^{\phantom{]\dagger}}+(\lambda_{k}/\omega_{k})\sigma^{z}$.
Here, the sums are over the fast oscillators only, and an unimportant
constant has been dropped. Notice that the difference between $a_{k}$
and $b_{k}$ simply represents a shift of the oscillator's rest position
of magnitude $x_{0k}=-\sigma^{z}\lambda_{k}\sqrt{2/(m_{k}\omega_{k}^{3})}$.
(Recall that $x_{k}\sqrt{2m_{k}\omega_{k}}\equiv a_{k}^{\dagger}+a_{k}^{\phantom{\dagger}}$,
with $m_{k}$ being the mass of the $k$-th oscillator.) Using the
translation operator, the shifted ground state of oscillator $k$
can be represented as 
\begin{equation}
\left|0_{k,\sigma^{z}}\right\rangle =e^{ix_{0k}p_{k}}\left|0_{k}\right\rangle \equiv\exp\left[\sigma^{z}\frac{\lambda_{k}}{\omega_{k}}\left(a_{k}^{\dagger}-a_{k}^{\phantom{\dagger}}\right)\right]\left|0_{k}\right\rangle ,
\end{equation}
 where $\left|0_{k}\right\rangle $ denotes the ground state of the
unperturbed oscillator. Finally, the ground state of $H_{0}$ is doubly
degenerate with eigenvectors 
\begin{equation}
\left|\Psi_{\pm}\right\rangle =\frac{1}{\sqrt{2}}\left(\left|+\right\rangle \prod_{\omega_{k}>ph}\left|0_{k,+}\right\rangle \pm\left|-\right\rangle \prod_{\omega_{k}>ph}\left|0_{k,-}\right\rangle \right).
\end{equation}
 The transverse-field term in the Hamiltonian, $H_{1}=-h\sigma^{x}$,
lifts this degeneracy. The tunnel splitting, i.e., the tunneling rate
of the spin $\sigma$ between $\left|+\right\rangle $ and $\left|-\right\rangle $,
is given by the energy difference $E_{+}-E_{-}=2h^{\prime}=\left\langle \Psi_{+}\left|H_{1}\right|\Psi_{+}\right\rangle -\left\langle \Psi_{-}\left|H_{1}\right|\Psi_{-}\right\rangle =2h\prod_{\omega_{k}>ph}\left\langle 0_{k,+}|0_{k,-}\right\rangle $.
Therefore, we can replace the spin coupled to the fast oscillators
by an effective free spin with a renormalized transverse field 
\begin{eqnarray}
h^{\prime} & = & h\prod_{\omega_{k}>ph}\left\langle 0_{k,+}|0_{k,-}\right\rangle =h\exp\left[-2\sum_{\omega_{k}>ph}(\lambda_{k}^{2}/\omega_{k}^{2})\right]\nonumber \\
 & = & h\exp\left[-\frac{2}{\pi}\int_{ph}^{\omega_{c}}\frac{{\cal E}\left(\omega\right)}{\omega^{2}}{\rm d}\omega\right]\label{eq:h-prime}
\end{eqnarray}
 which depends exponentially on the spectral density (\ref{eq:spectral-function}).
Now, as long as $h^{\prime}$ remains below the cutoff energy $\omega_{c}^{\prime}=ph$
of the remaining oscillators, this procedure can be iterated until
convergence.

The qualitative features of this adiabatic renormalization procedure
depend on the character of the dissipation.\cite{leggett-etal-rmp87}
For super-Ohmic dissipation ($s>1$), $h^{\prime}$ always converges
to a finite value $h^{*}>0$, regardless the dissipation strength
$\alpha$. Thus, the spin effectively \emph{decouples} from the bath
and remains tunneling in the presence of dissipation. In contrast,
$h^{*}$ vanishes for sub-Ohmic dissipation ($0<s<1$) implying that
the damping localizes the spin (at least in the limit of small $h$).
In the Ohmic case, the behavior depends on the dissipation strength
$\alpha$. As long as $\alpha<1$, the transverse field converges
to a nonzero value $h^{*}=h(ph/\omega_{c})^{\alpha/(1-\alpha)}$ while
it converges to $h^{*}=0$ for $\alpha>1$. Based on the heuristic
arguments of Ref.\ \onlinecite{vojta-prl03}, we therefore expect
a smeared quantum phase transition in the sub-Ohmic and Ohmic cases.

Let us now use the adiabatic renormalization of the oscillators within
the strong-disorder renormalization-group scheme of the full dissipative
random transverse-field Ising chain (\ref{eq:H}), as outlined in
Sec.\ \ref{subsec:SDRG-diss}. Our intent is to lower the renormalization-group
energy scale from $\Omega$ to $\Omega-{\rm d}\Omega$. To do so,
we need to integrate out all oscillators in the frequency range between
$p(\Omega-{\rm d}\Omega)$ and $p\Omega$. Because the interaction
terms in (\ref{eq:H}) are diagonal in the $\sigma^{z}$ basis, they
can simply be incorporated in $H_{0}$. Thus, when integrating out
the high-frequency oscillators in all baths, the transverse fields
renormalize according to 
\begin{alignat}{1}
h_{i}^{\prime} & =h_{i}\exp\left[-\alpha_{i}\omega_{I}^{1-s}\int_{p\Omega-p{\rm d}\Omega}^{p\Omega}{\rm d}\omega~\omega^{s-2}\right]\nonumber \\
 & =h_{i}\left[1-\alpha_{i}\left(\frac{\Omega}{\Omega_{I}}\right)^{s-1}\frac{{\rm d}\Omega}{\Omega}\right].\label{eq:new-h-adiabatic}
\end{alignat}
 Here, we have used the relation $\omega_{I}=p\Omega_{I}$ between
the bare renormalization group energy scale $\Omega_{I}$ and the
bare oscillator cutoff frequency $\omega_{I}$. Notice that the renormalized
fields $h_{i}^{\prime}$ do not depend on the arbitrary constant $p\gg1$.

Equation (\ref{eq:new-h-adiabatic}) reveals a crucial difference
between the super-Ohmic, Ohmic, and sub-Ohmic cases. In the super-Ohmic
case, the term renormalizing the transverse field is suppressed by
a factor $(\Omega/\Omega_{I})^{s-1}$, which vanishes as the energy
scale $\Omega$ goes to zero. This suggests that super-Ohmic dissipation
becomes unimportant at sufficiently low-energy scales. Our calculations
in the following will show that this is indeed the case. In contrast,
the term renormalizing the transverse field remains finite or even
diverges with $\Omega\to0$ in the Ohmic and sub-Ohmic cases, respectively.

\subsection{Decimating a site}

Let us now focus on integrating out a site, say site 2, because its
transverse field $h_{2}$ is at the renormalization-group energy scale
$\Omega$. It is important to realize that if $h_{2}=\Omega$, the
field actually has reached its fully converged value $h_{2}^{*}$
with respect to the adiabatic renormalization discussed in the last
subsection (because all remaining oscillators have frequencies below
$ph_{2}$). Hence, the spin $\sigma_{2}$ has decoupled from its bath,
and decimation of the site proceeds as in the undamped case.\ \cite{fisher95}
We consider $H_{0}=-h_{2}\sigma_{2}^{x}$ as the unperturbed Hamiltonian,
while the interactions of $\sigma_{2}$ with its neighbors are treated
perturbatively: $H_{1}=-(J_{1}\sigma_{1}^{z}+J_{2}\sigma_{3}^{z})\sigma_{2}^{z}$.
In second order of perturbation theory, the low-lying spectrum of
the cluster of sites 1, 2, and 3 can be represented by $H_{{\rm eff}}=-\tilde{J}\sigma_{1}^{z}\sigma_{3}^{z}$
with 
\begin{equation}
\tilde{J}=\frac{J_{1}J_{2}}{h_{2}}.\label{eq:J-tilde}
\end{equation}
 This means that the strongly fluctuating spin $\sigma_{2}$ is removed
from the chain, while the neighboring spins are coupled via a renormalized
weaker bond $\tilde{J}$.

\subsection{Decimating an interaction \label{subsec:interaction}}

Finally, the last possible decimation happens when an interaction,
say $J_{2}$, is at the renormalization group energy scale $\Omega$.
In this case, we consider the cluster of sites 2 and 3 as well as
its coupling to the neighboring sites 1 and 4. The unperturbed part
of the Hamiltonian reads as $H_{0}=-J_{2}\sigma_{2}^{z}\sigma_{3}^{z}$,
while $H_{1x}=-h_{2}\sigma_{2}^{x}-h_{3}\sigma_{3}^{x}$ and $H_{1z}=-J_{1}\sigma_{1}^{z}\sigma_{2}^{z}-J_{3}\sigma_{3}^{z}\sigma_{4}^{z}$
as well as $H_{1{\rm diss}}=\sum_{i=2}^{3}\sigma_{i}^{z}\sum_{k}\left[\lambda_{k,i}\left(a_{k,i}^{\dagger}+a_{k,i}^{\phantom{]\dagger}}\right)+\omega_{k,i}\left(a_{k,i}^{\dagger}a_{k,i}^{\phantom{]\dagger}}+1/2\right)\right]$
are treated as perturbations. As in the undamped case, the low-energy
states are those in which $\sigma_{2}^{z}$ and $\sigma_{3}^{z}$
are parallel. Therefore, the cluster of sites 2 and 3 can be replaced
by a single effective spin $\tilde{\sigma}$, the states $\left|+\right\rangle $
and $\left|-\right\rangle $ of which represent $\left|\sigma_{2}\sigma_{3}\right\rangle =\left|++\right\rangle $
and $\left|\sigma_{2}\sigma_{3}\right\rangle =\left|--\right\rangle $,
respectively. Treating the perturbations to lowest non-vanishing order
(first order of perturbation theory for $H_{1z}$ and $H_{1{\rm diss}}$
but second order for $H_{1x}$), yields the effective Hamiltonian
\begin{align}
H_{{\rm eff}}= & -J_{1}\sigma_{1}^{z}\tilde{\sigma}^{z}-J_{3}\tilde{\sigma}^{z}\sigma_{4}^{z}-\tilde{h}\tilde{\sigma}^{x}+\nonumber \\
 & \tilde{\sigma}^{z}\sum_{k}\tilde{\lambda}_{k}\left(a_{k}^{\dagger}+a_{k}^{\phantom{\dagger}}\right)+\omega_{k}\left(a_{k}^{\dagger}a_{k}^{\phantom{]\dagger}}+\frac{1}{2}\right)~.~\label{eq:Heff-spin-boson}
\end{align}
 The renormalized transverse field $\tilde{h}$ is smaller than either
of the original ones and given by 
\begin{equation}
\tilde{h}=\frac{h_{2}h_{3}}{J_{2}}~.\label{eq:h-tilde}
\end{equation}
 The magnetic moment associated with the renormalized spin $\tilde{\sigma}$
is the sum of the moments of $\sigma_{2}^{z}$ and $\sigma_{3}^{z}$,
\begin{equation}
\tilde{\mu}=\mu_{2}+\mu_{3},~\label{eq:mu-tilde}
\end{equation}
 while the dissipative bath coupled to $\tilde{\sigma}$ has spectral
density 
\begin{equation}
\tilde{{\cal E}}(\omega)=\frac{\pi}{2}\tilde{\alpha}\omega_{I}^{1-s}\omega^{s}~\mbox{ with }\tilde{\alpha}=\alpha_{2}+\alpha_{3}.\label{eq:alpha-tilde}
\end{equation}
 This means the renormalized dissipation strength of a cluster increases
with its magnetic moment. We can thus parametrize the local dissipation
strength $\alpha_{i}$ in terms of the magnetic moment $\mu_{i}$
as 
\begin{equation}
\alpha_{i}=\alpha\mu_{i}.\label{eq:alpha-mu}
\end{equation}
 This has the advantage that we do not have to separately keep track
of the probability distribution of the dissipation strengths during
the renormalization-group flow.

We note in passing that one can also include $H_{1{\rm diss}}$ in
the unperturbed Hamiltonian, as was done in Ref.\ \onlinecite{scherh-rieger-jsm08}.
This leads to more complicated algebra but results in the same recursions
(\ref{eq:Heff-spin-boson})--(\ref{eq:alpha-tilde}) at the approximation
level considered here.

\subsection{Summary of the decimation procedure}

\begin{figure}[t]
\centering{}\psfrag{(a)}{(a) $\Omega = \omega_c/p$}
\psfrag{(b)}{(b.i) $\Omega = h_2$}
\psfrag{(c)}{(b.ii) $\Omega = J_2$}
 \includegraphics[clip,width=1\columnwidth]{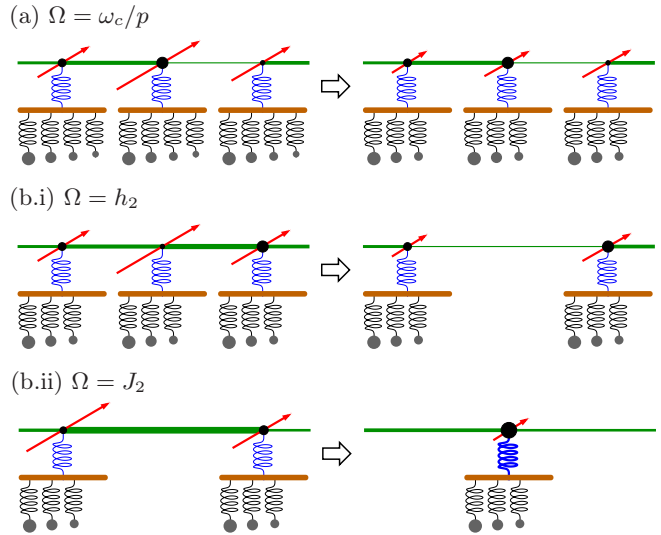} \caption{(Color online) Decimation scheme. The magnitude of the transverse
fields, interactions and magnetic moments are depicted by the length
of the arrows, the thickness of the lines, and the size of the dots,
respectively. The bosonic baths are depicted by the array of springs
and oscillators in which lighter (small) oscillators are faster. Finally,
the magnitude of the coupling to the bath (or the local dissipation
strength) is depicted by the thickness of the spring connecting the
spin to the bath. (a) \emph{Global} adiabatic renormalization of the
fast oscillators, (b.i) decimation of a strong \emph{local} transverse
field, and (b.ii) decimation of a strong \emph{local} interaction.
\label{fig:decimations}}
\end{figure}

The entire decimation procedure resulting from the steps outlined
in Secs.\ \ref{subsec:adiabatic} to \ref{subsec:interaction} can
be summarized as follows (see Fig.\ \ref{fig:decimations} for a
schematic illustration).

In lowering the renormalization group energy scale from $\Omega$
to $\Omega-{\rm d}\Omega$, we first integrate out all oscillators
with frequencies in the range $[p\Omega-p{\rm d}\Omega,p\Omega]$
in adiabatic renormalization. This modifies \emph{all} transverse
fields according to (\ref{eq:new-h-adiabatic}) {[}see Fig.\ \hyperref[fig:decimations]{\ref{fig:decimations}(a)}{]}.

After this global decimation, we integrate out all transverse fields
and interactions in the energy range $[\Omega-{\rm d}\Omega,\Omega]$.
(As these decimations are local they do not interfere with each other.)
In the case of decimating a transverse field, the corresponding site
and bath are removed from the system, yielding an effective coupling
between the neighboring spins given by (\ref{eq:J-tilde}) {[}see
Fig.\ \hyperref[fig:decimations]{\ref{fig:decimations}(b.i)}{]}.
When integrating out an interaction, the corresponding spins are replaced
by a single effective one. The transverse field acting on this spin
and its effective magnetic moment are given by Eqs.\ (\ref{eq:h-tilde})
and (\ref{eq:mu-tilde}). The spectral function of the bosonic bath
coupling to the effective spin has the same exponent $s$ as the original
baths, but the dissipation strengths increases according to (\ref{eq:alpha-tilde})
{[}see Fig.\ \hyperref[fig:decimations]{\ref{fig:decimations}(b.ii)}{]}.
This last renormalization is taken into account automatically if one
defines the local $\alpha_{i}$ as in Eq.\ (\ref{eq:alpha-mu}).

\section{Flow equations \label{sec:Flow-equations}}

The strong-disorder renormalization group consists in iterating the
decimation steps of Sec.\ \ref{sec:RG-recursion-relation}, and so
reducing the renormalization-group energy scale $\Omega=\max\{J_{i},h_{i},\omega_{c}/p\}$
to zero. In this process, the probability distributions of the interactions,
transverse fields, and magnetic moments change. In this section, we
derive the renormalization-group flow equations for these distributions.
In contrast to the dissipationless case, where the magnetic moment
does not enter the recursions for the transverse fields and interactions,\ \cite{fisher92,fisher95}
our recursion relation (\ref{eq:new-h-adiabatic}) couples the magnetic
moments $\mu$ and transverse fields $h$ (via $\alpha_{i}=\mu_{i}\alpha$).
In addition to the distribution $P(J;\Omega)$ of the interactions,
we therefore need to consider the \emph{joint} distribution of transverse
fields and magnetic moments $R(h,\mu;\Omega)$. The last argument
of these distributions makes explicit the dependence on the renormalization-group
energy scale $\Omega$. Occasionally, we will also use the reduced
distribution of the transverse fields, $R_{h}(h;\Omega)=\int_{0}^{\infty}R(h,\mu;\Omega){\rm d}\mu$.

\subsection{Flow of $P(J;\Omega)$ \label{subsec:P(J)}}

\begin{figure}[b]
\centering{}\psfrag{a}{(a) Before}
\psfrag{b}{(b) After}
\psfrag{0}{$0$}
\psfrag{mi}{$\mu$}
\psfrag{dO}{${\rm d}\Omega$}
\psfrag{h}{$h$}
\psfrag{hp}{$h^\prime$}
\psfrag{mic}{$\mu_c$}
\psfrag{miOalfa}{$\, \mu \!=\!\! \frac{\Omega - h^\prime}{\alpha \left( \frac{\Omega}{\Omega_I} \right)^{s-1} {\rm d}\Omega}$}
\includegraphics[clip,width=1\columnwidth]{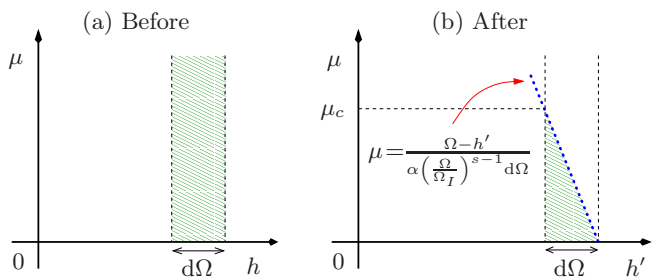} \caption{(Color online) Schematic of the possible transverse-field decimations
in the $h$ \emph{vs}. $\mu$ plane. (a) Before the integration of
the fast oscillators, the hatched area represents all the potential
fields to be decimated. (b) After integrating out the oscillators,
some of these fields have been shifted below $\Omega-d\Omega$ and
are not eligible to be decimated anymore, only those in the triangular
shaded area still are. The dashed line $h=\Omega$ maps into the dotted
line $\mu=(\Omega-h^{\prime})/(\alpha(\Omega/\Omega_{I})^{s-1}{\rm d}\Omega)$
after the adiabatic-renormalization step (\ref{eq:new-h-adiabatic}).
\label{fig:h-integration}}
\end{figure}

The distribution of the interactions $J$ only changes when a site
(a strong transverse field) is decimated. Upon reducing the energy
scale from $\Omega$ to $\Omega-{\rm d}\Omega$, $P(J;\Omega)$ thus
transforms almost as in the undamped case\ \cite{fisher95}: 
\begin{align}
 & P(J;\Omega-{\rm d}\Omega)\left[1-{\rm d}\Omega\left(P(\Omega;\Omega)+R_{h^{\prime}}^{\prime}(\Omega;\Omega)\right)\right]=\nonumber \\
 & P(J;\Omega)-{\rm d}\Omega R_{h^{\prime}}^{\prime}(\Omega;\Omega)\int{\rm d}J_{1}{\rm d}J_{2}P(J_{1};\Omega)P(J_{2};\Omega)\times\nonumber \\
 & \left[\delta\left(J-J_{1}\right)+\delta\left(J-J_{2}\right)-\delta\left(J-\frac{J_{1}J_{2}}{\Omega}\right)\right],\label{eq:flow-P}
\end{align}
 where $R_{h^{\prime}}^{\prime}(h^{\prime};\Omega)$ is the distribution
of transverse fields \emph{after} integrating out the fast bath oscillators
in the energy interval $[p\Omega-pd\Omega,p\Omega]$ according to
Eq.\ (\ref{eq:new-h-adiabatic}). The term between the brackets on
the left-hand side of (\ref{eq:flow-P}) guarantees the normalization
of $P(J;\Omega-{\rm d}\Omega)$ (decimating either a site or an interaction
each reduces the number of interactions in the system by one). The
second term on the right-hand side implements the decimation of a
site (transverse field), which happens with probability ${\rm d}\Omega R_{h^{\prime}}^{\prime}(\Omega;\Omega)$.
The first and second delta functions in this term remove the neighboring
couplings while the third one insert the renormalized one, as given
by (\ref{eq:J-tilde}). The integral over $J_{1}$ and $J_{2}$, weighted
by $P(J_{1};\Omega)P(J_{2};\Omega)$, sums over all possible values
of neighboring couplings.

The only difference to the undamped case stems from the fact that
the decimation of a site is determined by the distribution $R_{h^{\prime}}^{\prime}(h^{\prime};\Omega)$
of the transverse fields \emph{after} integrating out the fast bath
oscillators rather than the distribution of the unrenormalized fields
$R_{h}(h;\Omega)$. Using (\ref{eq:new-h-adiabatic}), the probability
$R_{h^{\prime}}^{\prime}(\Omega;\Omega){\rm d}\Omega$ of having a
decimation of a site can be found as 
\begin{eqnarray}
R_{h^{\prime}}^{\prime}(\Omega;\Omega){\rm d}\Omega=\int_{0}^{\infty}{\rm d}\mu\int_{\Omega-{\rm d}\Omega}^{\Omega}{\rm d}h^{\prime}R^{\prime}(h^{\prime},\mu;\Omega)\nonumber \\
=\int_{0}^{\infty}{\rm d}\mu\int_{\Omega-{\rm d}\Omega+\alpha\mu\left(\Omega/\Omega_{I}\right)^{s-1}{\rm d}\Omega}^{\Omega}{\rm d}hR(h,\mu;\Omega)\nonumber \\
=\left[1-\alpha\overline{\mu}_{\Omega}\left(\Omega/\Omega_{I}\right)^{s-1}\right]R_{h}(\Omega;\Omega){\rm d}\Omega,\label{eq:prob-dec-h}
\end{eqnarray}
 where $\overline{\mu}_{\Omega}=\int_{0}^{\infty}\mu R(\Omega,\mu;\Omega){\rm d}\mu/\int_{0}^{\infty}R(\Omega,\mu;\Omega){\rm d}\mu$
is the mean magnetic moment to be decimated at the energy scale $\Omega$.
Equation\ (\ref{eq:prob-dec-h}) can be understood as follows (see
Fig.\ \ref{fig:h-integration} for an illustration). Before integrating
out the fast oscillators, the potential fields to be decimated are
those in the range $[\Omega-{\rm d}\Omega,\Omega]$ highlighted in
Fig.\ \hyperref[fig:h-integration]{\ref{fig:h-integration}(a)}.
However, after eliminating these oscillators, some of those fields
are renormalized downward out of decimating region. Only the fields
in the hatched area of Fig.\ \hyperref[fig:h-integration]{\ref{fig:h-integration}(b)}
may still be decimated.

Equation (\ref{eq:prob-dec-h}) implies that any cluster with magnetic
moment larger than $\mu_{c}(\Omega)=1/(\alpha(\Omega/\Omega_{I})^{s-1})$
will never be decimated. For Ohmic dissipation, $s=1$, this means
that there is an energy-independent cutoff cluster size $\mu_{c}=1/\alpha$
(and thus a finite length scale). All clusters with $\mu>\mu_{c}$
will become frozen as their transverse fields $h$ iterate to zero
with $\Omega\rightarrow0$. This will lead to the smearing of the
quantum phase transition. In contrast, for super-Ohmic dissipation,
$s>1$, the cutoff cluster size $\mu_{c}(\Omega)$ diverges with $\Omega\rightarrow0$,
which reflects the fact that super-Ohmic dissipation cannot prevent
the tunneling of even the largest clusters.

\subsection{Flow of $R(h,\mu;\Omega)$ \label{subsec:R(h)}}

The joint distribution $R(h,\mu;\Omega)$ of transverse fields and
magnetic moments changes when an interaction is decimated. It also
changes in response to the adiabatic renormalization of the fast oscillators.
As a result, the flow equation of $R(h,\mu;\Omega)$ is more complicated
than in the dissipationless case:\ \cite{fisher95} \begin{widetext}
\begin{align}
R\left(h,\mu;\Omega-{\rm d}\Omega\right) & \left[1-{\rm d}\Omega\left(P(\Omega;\Omega)+R_{h^{\prime}}^{\prime}(\Omega;\Omega)\right)\right]=\nonumber \\
 & R\left(h,\mu;\Omega\right)-{\rm d}\Omega P(\Omega;\Omega)\int{\rm d}h_{2}{\rm d}\mu_{2}{\rm d}h_{3}{\rm d}\mu_{3}R(h_{2},\mu_{2;\Omega})R(h_{3},\mu_{3};\Omega)\nonumber \\
 & \qquad\times\left[\delta\left(h-h_{2}\right)\delta\left(\mu-\mu_{2}\right)+\delta\left(h-h_{3}\right)\delta\left(\mu-\mu_{3}\right)-\delta\left(h-\frac{h_{2}h_{3}}{\Omega}\right)\delta\left(\mu-\mu_{2}-\mu_{3}\right)\right]\nonumber \\
 & -\int{\rm d}h_{2}{\rm d}\mu_{2}R(h_{2},\mu_{2};\Omega)\left[\delta\left(h-h_{2}\right)\delta\left(\mu-\mu_{2}\right)-\delta\left(h-h_{2}^{\prime}\right)\delta\left(\mu-\mu_{2}\right)\right],\label{eq:flow-R}
\end{align}
 where $h_{2}^{\prime}$ is given by Eq.~(\ref{eq:new-h-adiabatic}).
The second term on the right-hand side of (\ref{eq:flow-R}) implements
the change in the fields and magnetic moments when an interaction
is decimated. The third term implements the adiabatic renormalization
due the decimation of the fast bath oscillators. The normalization
term {[}between the brackets on the left-hand side of (\ref{eq:flow-R}){]}
takes the same form as in (\ref{eq:flow-P}) because decimating either
a site or an interaction each reduces the number of sites (i.e., transverse
fields) in the system by one.

After some tedious but straightforward algebra, Eqs.\ (\ref{eq:flow-P})
and (\ref{eq:flow-R}) simplify to 
\begin{align}
-\frac{\partial}{\partial\Omega}P(J;\Omega)= & \left[P(\Omega;\Omega)-\left(1-\alpha\overline{\mu}_{\Omega}\left(\frac{\Omega}{\Omega_{I}}\right)^{s-1}\right)R_{h}(\Omega;\Omega)\right]P(J;\Omega)\nonumber \\
 & +\left(1-\alpha\overline{\mu}_{\Omega}\left(\frac{\Omega}{\Omega_{I}}\right)^{s-1}\right)R_{h}(\Omega;\Omega)\int{\rm d}J_{1}{\rm d}J_{2}P(J_{1};\Omega)P(J_{2};\Omega)\delta\left(J-\frac{J_{1}J_{2}}{\Omega}\right),\label{eq:flow-P-final}\\
-\frac{\partial}{\partial\Omega}R(h,\mu;\Omega)= & \left[\left(1-\alpha\overline{\mu}_{\Omega}\left(\frac{\Omega}{\Omega_{I}}\right)^{s-1}\right)R_{h}(\Omega;\Omega)-P(\Omega;\Omega)\right]R(h,\mu;\Omega)+\frac{\alpha\mu}{\Omega}\left(\frac{\Omega}{\Omega_{I}}\right)^{s-1}\left(1+h\frac{\partial}{\partial h}\right)R(h,\mu;\Omega)\nonumber \\
 & +P(\Omega;\Omega)\int{\rm d}h_{2}{\rm d}\mu_{2}{\rm d}h_{3}{\rm d}\mu_{3}R(h_{2},\mu_{2};\Omega)R(h_{3},\mu_{3};\Omega)\delta\left(h-\frac{h_{2}h_{3}}{\Omega}\right)\delta\left(\mu-\mu_{2}-\mu_{3}\right).\label{eq:flow-R-final}
\end{align}
\end{widetext} All terms proportional to $\alpha$ are due to dissipation.
Those multiplying $\overline{\mu}_{\Omega}$ correspond to a reduced
probability of decimating a site because the transverse fields are
renormalized downward by the dissipative baths. The additional term
(the second one on the right-hand side) in Eq.\ (\ref{eq:flow-R-final})
corresponds to the global change of all transverse fields when the
oscillators are integrated out. The flow equations of the undamped
case\ \cite{fisher92,fisher95} are recovered as expected if we set
$\alpha=0$.

Moreover, in the undamped case, the magnetic moments in (\ref{eq:flow-R-final})
can be integrated out, resulting in a flow equations for the reduced
distribution $R_{h}(h;\Omega)$ of the transverse fields. This is
no longer the case for $\alpha\neq0$, and the fixed-point distributions
for fields and magnetic moments have to be obtained jointly.

\section{Formal solution of the flow equations \label{sec:Formal-solution}}

In this section, we solve the renormalization-group flow equations
(\ref{eq:flow-P-final}) and (\ref{eq:flow-R-final}). To this end,
we first change to logarithmic energy variables. We then perform a
general rescaling transformation which allows us to identify fixed-point
distributions of the flow equations.

\subsection{Logarithmic variables}

We introduce a logarithmic measure of the renormalization-group energy
scale, 
\begin{equation}
\Gamma=\ln\left(\Omega_{I}/\Omega\right)\label{eq:log-energy-scale}
\end{equation}
 as well as the logarithmic variables 
\begin{equation}
\zeta=\ln\left(\Omega/J\right)~\mbox{ and }~\beta=\ln\left(\Omega/h\right),\label{eq:log-variables}
\end{equation}
 describing the interactions and transverse fields, respectively.\ \cite{fisher92,fisher95}
This is advantageous because the multiplicative recursion relations
(\ref{eq:J-tilde}) and (\ref{eq:h-tilde}) turn into additive ones:
$\tilde{\zeta}=\zeta_{1}+\zeta_{2}$ and $\tilde{\beta}=\beta_{2}+\beta_{3}$.

The probability distributions $\pi$ and $\rho$ of the logarithmic
variables can be defined as $\pi(\zeta;\Gamma){\rm d}\zeta=P(J;\Omega){\rm d}J$,
$\rho(\beta,\mu;\Gamma){\rm d}\beta{\rm d}\mu=R(h,\mu;\Omega){\rm d}h{\rm d}\mu$.
Rewriting the flow equations (\ref{eq:flow-P-final}) and (\ref{eq:flow-R-final})
in terms of these distributions yields 
\begin{align}
\frac{\partial\pi}{\partial\Gamma}= & \frac{\partial\pi}{\partial\zeta}+\left[\pi_{0}-\left(1-\alpha\overline{\mu}_{0}e^{-(s-1)\Gamma}\right)\rho_{\beta,0}\right]\pi(\zeta)\nonumber \\
 & +\left(1-\alpha\overline{\mu}_{0}e^{-(s-1)\Gamma}\right)\rho_{\beta,0}\pi\stackrel{\zeta}{\otimes}\pi,\label{eq:flow-pi}\\
\frac{\partial\rho}{\partial\Gamma}= & \frac{\partial\rho}{\partial\beta}+\left[\left(1-\alpha\overline{\mu}_{0}e^{-(s-1)\Gamma}\right)\rho_{\beta,0}-\pi_{0}\right]\rho(\beta,\mu)\nonumber \\
 & -\alpha\mu e^{-(s-1)\Gamma}\frac{\partial\rho}{\partial\beta}+\pi_{0}\rho\stackrel{\beta,\mu}{\otimes}\rho,\label{eq:flow-rho}
\end{align}
 where $\overline{\mu}_{0}\equiv\overline{\mu}_{\beta=0}=\overline{\mu}_{\Omega}$
is the mean magnetic moment to be decimated. $\pi_{0}=\pi(0;\Gamma)$
determines the probability of decimating an interaction while $\rho_{\beta,0}=\int\rho(0,\mu;\Gamma){\rm d}\mu$
determines the probability for decimating a transverse field. The
symbol $\otimes$ denotes the convolution, $\pi\stackrel{\zeta}{\otimes}\pi=\int{\rm d}\zeta_{1}{\rm d}\zeta_{2}\pi(\zeta_{1})\pi(\zeta_{2})\delta\left(\zeta-\zeta_{1}-\zeta_{2}\right)$.

The origin of each term in (\ref{eq:flow-pi}) and (\ref{eq:flow-rho})
can be easily tracked. The first term on the right-hand side of each
equation is due to the change of the variable $\zeta$ or $\beta$
when $\Gamma$ changes. The second terms ensure the normalizations
of the distributions. The last ones are responsible for implementing
the recursion relations (\ref{eq:J-tilde}) and (\ref{eq:h-tilde}),
and the third term in (\ref{eq:flow-rho}) implements the damping
of the transverse fields according (\ref{eq:new-h-adiabatic}).

\subsection{Rescaling\label{sub:rescaling}}

We now look for fixed-point solutions of the flow equations (\ref{eq:flow-pi})
and (\ref{eq:flow-rho}), i.e., for distributions $\pi$ and $\rho$
that are stationary under the renormalization-group flow. From the
solution of the dissipationless case,\ \cite{fisher92,fisher95}
it is known that such fixed points only emerge after the variables
$\zeta$ and $\beta$ are appropriately rescaled. Let us consider
the general transformations $\eta=\zeta/f_{\zeta}(\Gamma)$, $\theta=\beta/f_{\beta}(\Gamma)$,
and $\nu=\mu/f_{\mu}(\Gamma)$, where the scale factors $f_{\zeta,\beta,\mu}(\Gamma)$
are functions of the logarithmic energy cutoff $\Gamma$ only. Transforming
the flow equations to the distributions ${\cal P}(\eta)$ and ${\cal R}(\theta,\nu)$
of the rescaled variables yields 
\begin{align}
\frac{\partial{\cal P}}{\partial\Gamma}= & \frac{\dot{f}_{\zeta}}{f_{\zeta}}\left({\cal P}+\eta\frac{\partial{\cal P}}{\partial\eta}\right)+\frac{1}{f_{\zeta}}\left(\frac{\partial{\cal P}}{\partial\eta}+{\cal P}_{0}{\cal P}\right)\nonumber \\
 & +\frac{1}{f_{\beta}}\left(1-\alpha\overline{\nu}_{0}f_{\mu}e^{-(s-1)\Gamma}\right){\cal R}_{\theta,0}\left({\cal P}\stackrel{\eta}{\otimes}{\cal P}-{\cal P}\right),\label{eq:flow-rescale-P}\\
\frac{\partial{\cal R}}{\partial\Gamma}= & \frac{\dot{f}_{\beta}}{f_{\beta}}\left({\cal R}+\theta\frac{\partial{\cal R}}{\partial\theta}\right)+\frac{\dot{f}_{\mu}}{f_{\mu}}\left({\cal R}+\nu\frac{\partial{\cal R}}{\partial\nu}\right)\nonumber \\
 & +\frac{1}{f_{\beta}}\left[\frac{\partial{\cal R}}{\partial\theta}+\left(1-\alpha\overline{\nu}_{0}f_{\mu}e^{-(s-1)\Gamma}\right){\cal R}_{\theta,0}{\cal R}\right]\nonumber \\
 & -\alpha\nu\frac{f_{\mu}}{f_{\beta}}e^{-(s-1)\Gamma}\frac{\partial{\cal R}}{\partial\theta}+\frac{1}{f_{\zeta}}{\cal P}_{0}\left({\cal R}\stackrel{\theta,\nu}{\otimes}{\cal R}-{\cal R}\right),\label{eq:flow-rescale-R}
\end{align}
 where $\dot{f}={\rm d}f/{\rm d}\Gamma$, $\overline{\nu}_{0}\equiv\overline{\nu}_{\theta=0}$,
${\cal P}_{0}={\cal P}(0;\Gamma)$, and ${\cal R}_{\theta,0}=\int{\cal R}(0,\nu;\Gamma){\rm d}\nu$.

\subsection{Fixed-point distributions}

The low-energy solutions of the flow equations describe stable phases
or critical points. Before analyzing the equations in detail, we can
already identify two simple solutions that correspond to the conventional
paramagnetic and ferromagnetic phases. The key feature is a separation
of scales between the interactions and the transverse fields.

When all interactions $J$ are larger than all transverse fields $h$
(i.e., their distributions do not overlap), the renormalization-group
procedure will only decimate interactions until every site is included
in the infinite cluster formed. This happens at a finite energy scale,
viz., the minimum value of $J$ in the chain. The system is thus in
a conventional, homogenously ordered, ferromagnetic phase.

Analogously, if all interactions are smaller than all transverse fields,
only sites will be decimated under the renormalization group. Strictly,
the interactions need to be smaller than all \emph{fully renormalized}
transverse fields in the sense of the adiabatic renormalization of
Sec.\ \ref{subsec:adiabatic} (which also requires sufficiently weak
dissipation). This is the conventional paramagnetic phase. As above,
there is a finite characteristic energy, viz., the minimum value of
the (renormalized) transverse fields in the system.

In the absence of dissipation, these two conventional solutions are
gapped, with the energy gap given by the final renormalization-group
energy scale discussed above. In the presence of the dissipative baths,
a true energy gap does not exist, but the final renormalization-group
scale sets a characteristic energy for the magnetic excitations in
the system.

After these simple solutions, we now turn our attention to the nontrivial
ones at which the interactions $J$ and transverse fields $h$ compete
with each other. This means, the distributions of the interactions
and (renormalized) transverse fields overlap. In these cases, we expect
the characteristic renormalization-group energy scale to vanish, either
because the system is critical or due to a quantum-Griffiths mechanism.
We therefore take $\Gamma\rightarrow\infty$ and search for $\Gamma$-independent
solutions of the rescaled flow equations (\ref{eq:flow-rescale-P})
and (\ref{eq:flow-rescale-R}). In this analysis, we need to distinguish
the different types of dissipation.

\subsubsection{Review of the dissipationless case\label{sub:Dissipationless-case}}

In the dissipationless case,\ \cite{fisher95,igloi-det-z-PRB} the
analysis is simplified by integrating out the magnetic moments. Their
distribution can be found separately after the fixed-point distributions
${\cal P}^{*}(\eta)$ and ${\cal R}_{\theta}^{*}(\theta)$ have been
obtained.

There are three types of fixed points. (i) A critical solution for
which duality requires that $f_{\zeta}=f_{\beta}$. In order to have
a physical (normalizable and non-oscillatory) distribution ${\cal P}(\eta)$,
we conclude from (\ref{eq:flow-rescale-P}) that $\dot{f}_{\zeta}/f_{\zeta}$
must scale like $1/f_{\zeta}$ with $\Gamma$, implying $f_{\zeta}\sim\Gamma$.\ \cite{footnote1}
Further analysis (requiring that $\overline{\nu}_{0}>0$) leads to
$f_{\mu}\sim\Gamma^{\phi}$, with $\phi=(1+\sqrt{5})/2$ being the
golden mean. The fixed-point distributions of the rescaled interactions
and transverse fields are ${\cal P}^{*}(x)={\cal R}_{\theta}^{*}(x)=e^{-x}$.
There does not seem to be a simple closed-form expression for the
joint fixed-point distribution of fields and moments, ${\cal R}^{*}(\theta,\nu)$.

(ii) On the ordered side of the transition, there is a line of fixed
points parameterized by ${\cal P}_{0}$ describing the ferromagnetic
quantum Griffiths phase. The scale factors are $f_{\zeta}=1$ and
$f_{\beta}=f_{\mu}=\exp({\cal P}_{0}\Gamma)$. The fixed-point distributions
are given by ${\cal P}^{*}(\eta)={\cal P}_{0}e^{-{\cal P}_{0}\eta}$,
${\cal R}^{*}(\theta,\nu)={\cal R}_{\nu,0}{\cal R}_{\theta,0}e^{-{\cal R}_{\theta,0}\theta}\delta\left({\cal R}_{\theta,0}\theta-{\cal R}_{\nu,0}\nu\right)$,
with ${\cal R}_{\theta,0}$ and ${\cal R}_{\nu,0}$ being nonuniversal
constants (see Appendix \ref{sec:Flow-equation-FM}).

(iii) Finally, the third type of solution is a line of fixed points
on the disordered side of the transition. They are parameterized by
${\cal R}_{0}$ and describe the paramagnetic quantum Griffiths phase.
The scale factors are $f_{\beta}=1$ and $f_{\zeta}=\exp({\cal R}_{0}\Gamma)$.
The resulting fixed-point distributions for the interactions and transverse
fields read ${\cal P}^{*}(\eta)={\cal P}_{0}e^{-{\cal P}_{0}\eta}$
and ${\cal R}_{\theta}^{*}(\theta)={\cal R}_{0}e^{-{\cal R}_{0}\theta}$
with ${\cal P}_{0}$ a nonuniversal constant. The mean value of the
magnetic moments scales as $\overline{\mu}\sim\Gamma$,\ \cite{fisher95}
which suggests that $f_{\mu}=\Gamma$. However, there is no fixed-point
distribution $R_{\nu}^{*}(\nu)$.\ \cite{hoyos-unpublished} In fact,
if one assumes that ${\cal R}^{*}(\theta,\nu)$ exists, one arrives
at the incorrect conclusion that $\overline{\mu}={\rm const}$.\ \cite{igloi-det-z-PRB}

\subsubsection{Super-Ohmic dissipation\label{subsub:superohmic}}

To investigate the rescaled flow equations (\ref{eq:flow-rescale-P})
and (\ref{eq:flow-rescale-R}) in the super-Ohmic ($s>1$) case, we
first consider the stability of the dissipationless fixed points of
Sec.\ \ref{sub:Dissipationless-case} with respect to the dissipative
terms. In all dissipative terms in (\ref{eq:flow-rescale-P}) and
(\ref{eq:flow-rescale-R}), the dissipation strength $\alpha$ appears
in the combination $\alpha f_{\mu}e^{-(s-1)\Gamma}/f_{\beta}$. To
analyze the behavior of these terms close to the dissipationless fixed
points, we use the scale factors $f_{\mu}$ and $f_{\beta}$ discussed
in Sec.\ \ref{sub:Dissipationless-case}. For all three types of
dissipationless fixed points (critical, ferromagnetic Griffiths, and
paramagnetic Griffiths), we find that the dissipative terms are subleading
and vanish in the limit $\Gamma\to\infty$ for any $s>1$. This means
that the dissipationless fixed points are stable against weak super-Ohmic
damping. In other words, super-Ohmic dissipation is an irrelevant
perturbation (in the renormalization-group sense) of the dissipationless
random transverse-field Ising chain.

We have also checked that the dissipationless fixed-point solutions
are the only physical ones by tediously inspecting the flow equations
(\ref{eq:flow-rescale-P}) and (\ref{eq:flow-rescale-R}). If we assume
that the dissipative terms are the leading ones, we always arrive
at unphysical fixed-point distributions.

Instead of reproducing these calculations, we now present a more intuitive
argument. $\beta$ and $\mu$ are both renormalized when an interaction
is decimated, and both have the same additive recursion relations,
$\tilde{\beta}=\beta_{2}+\beta_{3}$ and $\tilde{\mu}=\mu_{2}+\mu_{3}$.
Therefore, one might expect that $f_{\mu}\sim f_{\beta}$. This is
not entirely true because $\beta$ depends explicitly on the value
of the renormalization-group scale $\Gamma$ and thus suffers an additional
downward shift of $-{\rm d}\Gamma$ as the renormalization-group scale
is increased from $\Gamma$ to $\Gamma+{\rm d}\Gamma$. As a result,
$f_{\mu}/f_{\beta}$ can diverge in the limit $\Gamma\to\infty$,
but never faster than $O(\Gamma)$. Moreover, in the presence of dissipation,
integrating out the oscillators via Eq.\ (\ref{eq:new-h-adiabatic})
leads to a downward renormalization of the transverse fields and thus
an increase of $\beta$. A fixed-point solution then requires a corresponding
increase in $f_{\beta}$. It is thus clear, that for any physical
fixed-point solution of the flow equations (\ref{eq:flow-rescale-P})
and (\ref{eq:flow-rescale-R}), the dissipative terms, which are proportional
to $\alpha f_{\mu}e^{-(s-1)\Gamma}/f_{\beta}$, remain exponentially
suppressed for any $s>1$.

How do we interpret these results physically? The corrections to the
effective transverse fields due to super-Ohmic damping become smaller
and smaller with decreasing renormalization-group energy scale $\Omega$
and increasing size (magnetic moment) of the clusters. This holds
both at criticality and in the surrounding quantum Griffiths phases.
The quantum phase transition of the super-Ohmic random transverse-field
Ising chain is thus identical to that of the dissipationless chain.
Even duality should be recovered at criticality. All these results
are in agreement with numerical work in Ref.\ \onlinecite{scherh-rieger-jsm08}.

\subsubsection{Ohmic dissipation\label{subsub:ohmic}}

In contrast to the super-Ohmic case, Ohmic dissipation is not an irrelevant
perturbation. The dissipative terms in the rescaled flow equations
(\ref{eq:flow-rescale-P}) and (\ref{eq:flow-rescale-R}) are not
exponentially suppressed and become important in \emph{all} of the
three regimes (critical, ferromagnetic Griffiths, paramagnetic Griffiths).
Hence, qualitatively different behavior is expected.

In this section, we show that the Ohmic dissipation destroys both
the paramagnetic quantum Griffiths phase and the quantum critical
point. Only a ferromagnetic ``Griffiths-type'' phase survives. Before
discussing the technical details, let us understand this result physically.
In a putative paramagnetic quantum Griffiths phase, there are arbitrarily
large ferromagnetic clusters embedded in the paramagnetic bulk. As
discussed at the end of Sec.\ \ref{subsec:P(J)}, clusters with magnetic
moments larger than $\mu_{c}=\alpha^{-1}$ can never be decimated
in the case of Ohmic dissipation. In the low-energy limit, they therefore
become frozen and develop true static magnetic order. These static
clusters will be aligned by any infinitesimally small interaction,
giving rise to long-range order. The system is thus an inhomogeneous
ferromagnet rather than a paramagnet.

Let us now see how the same result follows from the flow equations
(\ref{eq:flow-rescale-P}) and (\ref{eq:flow-rescale-R}). We start
by assuming that there is a paramagnetic quantum Griffiths phase.
In such a phase, the majority of renormalization-group steps would
be decimations of sites which renormalize the distribution ${\cal P}$
of the interactions. The convolution term ${\cal P}\stackrel{\eta}{\otimes}{\cal P}$
therefore has to be a leading term in the flow equation (\ref{eq:flow-rescale-P})
to describe the downward flow of the interactions under repeated decimations
of sites. Moreover, as $(1-\alpha\overline{\nu}_{0}f_{\mu}){\cal R}_{\theta,0}$
is the probability for decimating a transverse field, $1-\alpha\overline{\nu}_{0}f_{\mu}=c$
needs to remain nonzero and positive in the limit $\Gamma\to\infty$.
Thus, a physical solution from (\ref{eq:flow-rescale-P}) requires
that $f_{\zeta}$ diverges exponentially and $f_{\beta}$ is a constant
which we can set to 1. Inserting this into (\ref{eq:flow-rescale-R})
leads to the fixed-point solution ${\cal R}^{*}=A(\nu)\exp[-c{\cal R}_{\theta,0}^{*}\theta/(1-\alpha\nu f_{\mu})]$,
with $A(\nu)$ being a function of $\nu$ ensuring normalization and
${\cal P}^{*}={\cal P}_{0}\exp(-{\cal P}_{0}^{*}\eta)$, with $f_{\beta}=1$,
$f_{\mu}={\rm const}$, and $f_{\zeta}=\exp(c{\cal R}_{\theta,0}^{*}\Gamma)$.

However, this solution implies that there are no spin clusters with
moments bigger than $\mu_{c}=\nu_{c}f_{\mu}=1/\alpha$. Thus, this
solution cannot describe a quantum Griffiths phase in which, due to
statistical disorder fluctuations, arbitrarily large magnetic clusters
exist with small but nonzero probability. (Whenever the distributions
of interactions and transverse fields overlap, it is always possible
to find an arbitrarily large region in which all interactions are
bigger than all transverse fields.)

Using the same arguments, it also follows that a quantum critical
solution of the flow equations (\ref{eq:flow-rescale-P}) and (\ref{eq:flow-rescale-R})
(at which $f_{\mu}$ has to diverge to reflect the diverging correlation
length) can not exist.

After having seen that Ohmic dissipation destroys both the paramagnetic
quantum Griffiths phase and the quantum critical point (because no
physical fixed-point solution is possible when the convolution term
${\cal P}\stackrel{\eta}{\otimes}{\cal P}$ is present), we now consider
ferromagnetic solutions. On the ferromagnetic side of the transition,
almost all renormalization-group steps (at low energies) involve decimations
of interactions, while almost no sites are decimated. Therefore, the
convolution term ${\cal P}\stackrel{\eta}{\otimes}{\cal P}$ drops
out of the flow equation for $\Gamma\to\infty$. Moreover, the flow
of the interactions close to the fixed point should be analogous to
that of the undamped case because their decimations are not influenced
by the additional renormalization due to the bosonic baths. Therefore,
we use the ansatz $f_{\zeta}=1$ as in the undamped case. The flow
equation (\ref{eq:flow-rescale-P}) then yields 
\begin{equation}
{\cal P}^{*}(\eta)={\cal P}_{0}e^{-{\cal P}_{0}\eta},\label{eq:P-FP-ohmic}
\end{equation}
 where ${\cal P}_{0}$ is a nonuniversal constant which parameterizes
the line of ferromagnetic fixed points.

Let us now turn our attention to the joint distribution ${\cal R}^{*}(\theta,\nu)$
of transverse fields and magnetic moments at these fixed points. In
the dissipationless case (Sec.\ \ref{sub:Dissipationless-case})
fields and moments rescale in the same way, $f_{\mu}=f_{\beta}$.
In the presence of Ohmic dissipation, the transverse fields suffer
additional downward renormalizations due to Eq.\ (\ref{eq:new-h-adiabatic}),
which become more and more important as larger ferromagnetic clusters
are formed with increasing $\Gamma$. We thus expect that $f_{\mu}/f_{\beta}\rightarrow0$
with $\Gamma\rightarrow\infty$. Under this assumption, the flow equation
(\ref{eq:flow-rescale-R}) greatly simplifies, and all dissipative
terms drop out. This implies that the functional form of the fixed-point
distribution in terms of $\theta$ and $\nu$ is identical to that
of the undamped system in the ferromagnetic quantum Griffiths phase.
As in Ref.\ \onlinecite{fisher95}, we can integrate over either
$\nu$ or $\beta$ and analyze the resulting reduced distributions
${\cal R}_{\theta}^{*}$ and ${\cal R}_{\nu}^{*}$. This yields the
fixed-point distributions ${\cal R}_{\theta}^{*}={\cal R}_{\theta0}e^{-{\cal R}_{\theta0}\theta}$
and ${\cal R}_{\nu}^{*}={\cal R}_{\nu0}e^{-{\cal R}_{\nu0}\nu}$,
just as in the undamped case.

To fix the free parameters ${\cal R}_{\theta0}$ and ${\cal R}_{\nu0}$
and to find the joint distribution ${\cal R}^{*}(\theta,\nu)$, we
need to take into account the subleading term in the flow equation
(\ref{eq:flow-rescale-R}) close to the fixed point. In contrast to
the leading terms, the subleading ones do depend on the dissipation.
The details of this somewhat lengthy analysis are presented in Appendix\ \ref{sec:Flow-equation-FM}.
We find $f_{\mu}=e^{{\cal P}_{0}\Gamma}$ and $f_{\beta}=\Gamma e^{{\cal P}_{0}\Gamma}$
as well as ${\cal R}_{\nu0}/{\cal R}_{\theta0}=\alpha$. Moreover,
the rescaled magnetic moments and transverse fields become perfectly
correlated at the fixed point, such that the joint distribution reads
as 
\begin{equation}
{\cal R}^{*}(\beta,\nu)={\cal R}_{0}e^{-{\cal R}_{0}\nu}\delta\left(\theta-\alpha\nu\right).\label{eq:R-FP-ohmic}
\end{equation}
 The remaining nonuniversal constant ${\cal R}_{0}$ depends on the
initial conditions.

To summarize, instead of a quantum critical fixed point accompanied
by lines of quantum Griffiths fixed points on each side of the transition,
we only find a ferromagnetic solution that describes the physics for
all overlapping distributions of interactions and renormalized transverse
fields. To test the stability of this fixed point, we have computed-closed
form solutions of the entire renormalization-group flow for some particularly
simple initial distributions of transverse fields and interactions
(see Appendix\ \ref{sec:Flow-equation-FM}). In addition, we have
implemented the recursion relations numerically\ \cite{vojta-hoyos-procee-dissip,hoyos-unpublished}
and verified that the distributions flow toward our fixed point for
many different initial distributions.

\subsubsection{Sub-Ohmic dissipation\label{subsub:sub-Ohmic-flow}}

For the sub-Ohmic random transverse-field Ising chain, our strong-disorder
renormalization group method cannot be applied because the adiabatic
renormalization of the field does not capture all of the relevant
physics of the problem. In the sub-Ohmic case, the adiabatic renormalization
only works in the limit of weak transverse fields.\ \cite{leggett-etal-rmp87}
In this limit, the dissipation suppresses the tunneling for any dissipation
strength $\alpha$. This means, the transverse fields all renormalize
to zero for any $\alpha$. Thus, within the adiabatic renormalization
approach to the baths, the sub-Ohmic random transverse-field Ising
chain does not have a quantum phase transition, and it is always in
the ferromagnetic ground state.

However, it is known from more advanced techniques that the sub-Ohmic
spin-boson model with fixed weak dissipation does undergo a quantum
phase transition at some nonzero value of the transverse field (see,
e.g., Refs.\ \onlinecite{kehrein-mielke-pla96,bulla-tong-vojta-prl03}).
In our context, this implies that small clusters fluctuate but the
dynamics of sufficiently large clusters freezes (their transverse
fields vanish under renormalization) just like in the Ohmic case.
We will briefly discuss the resulting behavior in Sec.\ \ref{sec:Discussions}.

\section{Phase diagram and observables \label{sec:Phase-diagram}}

Using the fixed-point distributions found in Sec.\ \ref{sec:Formal-solution},
we can now determine the phase diagram and compute thermodynamic observables
close to the quantum phase transition.

\subsection{Phase diagram and crossovers}

Let us contrast the infinite-randomness critical-point scenario of
the dissipationless random transverse-field Ising chain\ \cite{fisher92,fisher95}
(which, according to Sec.\ \ref{subsub:superohmic}, also applies
to super-Ohmic damping) with the smeared transition scenario emerging
for Ohmic dissipation.

\begin{figure}
\centering{}\psfrag{T}{{\large $T$}}
\psfrag{Z}{{\large $0$}}
\psfrag{T*}{{\large $T^{*}$}}
\psfrag{T**}{{\large $T^{**}$}}
\psfrag{h}{{\large $h_{typ}$}}
\psfrag{a}{(a) Sharp (undamped or superOhmic damping)}
\psfrag{b}{(b) Smeared (Ohmic damping)}
\psfrag{c}{(c) Sharp ($\alpha=0$, $d>1$)}
\psfrag{d}{(d) Smeared ($\alpha \ne 0$, $d>1$)}
\psfrag{Griffiths}{{\large Griffiths}}
\psfrag{PM}{{\color{red} {\bf {\large PM}}}}
\psfrag{FM}{{\color{blue} {\bf {\large FM}}}}
\psfrag{QC}{{\color{OliveGreen} {\bf {\large QC}}}}
\psfrag{IRFP}{{\color{OliveGreen} {\bf {\large IRFP}}}}
\psfrag{SO}{{\large SO}}
\psfrag{SD}{{\large SD}}
\psfrag{WO}{{\large WO}}
\psfrag{WD}{{\large WD}}
\psfrag{IO}{{\large IO}}
\includegraphics[clip,width=0.8\columnwidth]{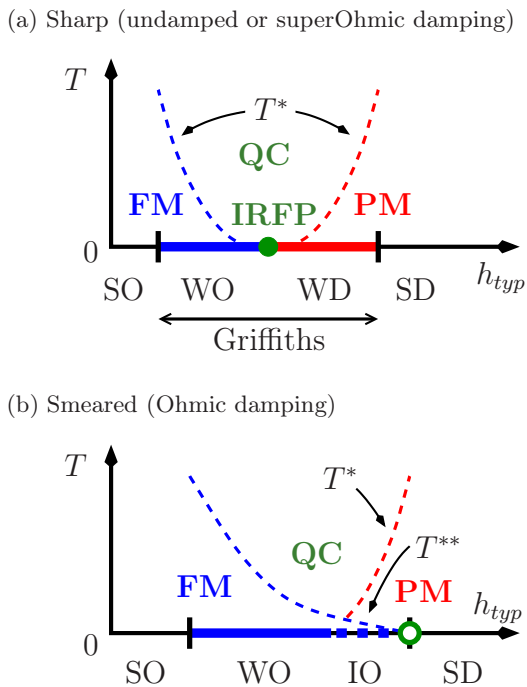} \caption{(Color online) Schematic phase diagrams in the transverse-field ($h_{{\rm typ}}$)
--- temperature ($T$) plane. (a) Infinite-randomness critical-point
scenario of the dissipationless and super-Ohmic chains. SO and SD
denote the strongly ordered and disordered conventional phases while
WO and WD denote the ordered and disordered quantum Griffiths phases.
(b) Smeared transition scenario for Ohmic dissipation ($0<\alpha\ll1)$,
with the inhomogeneously ordered (IO) ferromagnet replacing the WD
Griffiths phase. The open circle marks the end of the tail of the
smeared transition. The thin dashed lines represent finite-temperature
crossovers between a quantum critical (QC) regime and a quantum paramagnet
(PM) or ferromagnet (FM). The crossover temperatures $T^{\ast}$ and
$T^{\ast\ast}$ are discussed in the text. \label{fig:PD}}
\end{figure}

The phase diagram of the dissipationless case as function of temperature
$T$ and typical transverse field $h_{{\rm typ}}$ is sketched in
Fig.\ \hyperref[fig:PD]{\ref{fig:PD}(a)} (keeping the typical interaction
strength fixed).

At zero temperature, the ferromagnetic-paramagnetic quantum phase
transition is governed by a universal infinite-randomness fixed point
(IRFP). It is accompanied on both sides by the gapless paramagnetic
(weakly disordered) and ferromagnetic (weakly ordered) quantum Griffiths
phases. For higher fields (to the right of the paramagnetic Griffiths
phase), there is a conventional (strongly disordered) gapped quantum
paramagnet where quenched disorder effects are irrelevant. For sufficiently
weak fields (to the left of the ferromagnetic Griffiths phase), the
system is a conventional (strongly ordered) gapped ferromagnet where
quenched disorder is also irrelevant.

As ferromagnetic order in one dimension is destabilized at any nonzero
temperature, only crossovers will occur for $T>0$ {[}as depicted
by the dashed lines in Fig.\ \hyperref[fig:PD]{\ref{fig:PD}(a)}{]}.
Above the critical point, there is the quantum critical region which
is characterized by activated scaling, i.e., the dynamical exponent
is formally infinity. If the system is close to but not at the quantum
critical point, it undergoes a crossover to one of the quantum Griffiths
phases upon lowering the temperature. In these phases the dynamical
scaling is conventional, i.e., the dynamical exponent is finite, $z\sim1/|r|$,
where $r=h_{{\rm typ}}-h_{c}$ is the distance from criticality and
$h_{c}$ is the critical field. In the undamped case, duality requires
the critical point to occur when the typical interactions equal the
typical transverse fields, thus $h_{c}=J_{{\rm typ}}$. For super-Ohmic
dissipation, the fields are renormalized downward by the oscillators.
Using (\ref{eq:h-prime}) we find 
$h_{c}\approx J_{{\rm typ}}\exp\{\alpha/(s-1)\}$. The crossover temperature
$T^{*}$ is obtained from the activated dynamical scaling at the IRFP,
\begin{equation}
T^{*}\sim\exp(-{\rm const}/|r|^{-\nu\psi}),\label{eq:xover-undamped}
\end{equation}
 where $\nu=2$ and $\psi=1/2$ are the correlation length exponent
and tunneling exponent of the IRFP, respectively, and the constant
is of order unity.

Having reviewed the undamped case, let us now discuss the case of
Ohmic dissipation. For weak damping, $\alpha\ll1$, the renormalization
group flow at high energies (where all clusters are still small) will
be almost identical to that of the undamped system. Thus, the high
temperature part of the phase diagram shown in Fig.\ \hyperref[fig:PD]{\ref{fig:PD}(b)}
is analogous to its undamped counterpart. However, upon lowering the
energy scale (i.e., the temperature), the clusters grow. In the undamped
quantum Griffiths paramagnet discussed above their average magnetic
moment grows as $\overline{\mu}\sim r^{\nu\psi\left(1-\phi\right)}\ln(\Omega_{I}/T)$,
with $\phi=(1+\sqrt{5})/2$, and at the quantum critical point it
increases as $\overline{\mu}\sim[\ln(\Omega_{I}/T)]^{\phi}$.\ \cite{fisher95}
Therefore, it is clear that at some crossover temperature $T^{**}$,
the cluster moment reaches the critical value $\mu_{c}=1/\alpha$;
and dissipation becomes important. The value of $T^{**}$ can be estimated
from the condition $\overline{\mu}=1/\alpha$; this gives

\begin{equation}
T^{**}\sim\exp[-{\rm const}\times(1/\alpha)^{1/\phi}]\label{eq:xover-damped-hc}
\end{equation}
 very close the quantum critical point of the undamped system, and 

\begin{equation}
T^{**}\sim\exp(-{\rm const}\times r^{\nu\psi\left(\phi-1\right)}/\alpha)\label{eq:xover-damped}
\end{equation}
for larger transverse fields.

At temperatures below $T^{\ast\ast}$, the quantum dynamics of the
surviving clusters freezes completely (because their renormalized
transverse fields vanish). These clusters thus behave like large \emph{classical}
spins. At zero temperature, they can be aligned by any infinitesimally
weak interaction, leading to long-range order. Thus, the quantum Griffiths
paramagnet (and quantum critical point) are replaced by an inhomogeneously
ordered (IO) ferromagnetic phase {[}marked by the thick dashed line
in Fig.\ \hyperref[fig:PD]{\ref{fig:PD}(b)}. This phase is due to
\emph{local} formation of magnetic order rather than a collective
effect, which implies that the paramagnetic-ferromagnetic quantum
phase transition is smeared. It also means that the end point of the
tail of the smeared transition {[}marked by the open circle in Fig.\ \hyperref[fig:PD]{\ref{fig:PD}(b)}{]}
is \emph{not} a critical point.

\subsection{Observables}

We now turn our attention to observables. As shown in Sec.\ \ref{subsub:superohmic},
super-Ohmic dissipation is an irrelevant perturbation at the infinite-randomness
critical point of the dissipationless random transverse-field Ising
chain. Therefore, the low-energy behavior of observables in the super-Ohmic
case is identical to that in the dissipationless case, which is discussed
in detail in Ref.\ \onlinecite{fisher95} (see also Ref.\ \onlinecite{igloi-review}
for a review of additional results). This observation agrees with
numerical results of Schehr and Rieger.\ \cite{scherh-rieger-jsm08}

Here, we therefore only consider the case of Ohmic dissipation. Moreover,
we focus on the inhomogeneously ordered (IO) ferromagnetic phase (i.e.,
on the tail of the smeared transition), which is the novel feature
of the Ohmic chain. As we have seen in the analysis of the flow equations
in Sec.\ \ref{sec:Formal-solution}, weak dissipation does not change
the behavior of the other phases qualitatively; it only leads to quantitative
corrections to the undamped physics.

To characterize the extent to which the bare (initial) distributions
$P_{I}$ and $R_{I}$ of the interactions and transverse fields overlap,
we introduce the probability of an interaction $J$ being larger than
a renormalized transverse field, 
\begin{equation}
w=\int_{0}^{\infty}{\rm d}JP_{I}(J)\int_{0}^{J}{\rm d}h_{{\rm eff}}R_{I}(h_{{\rm eff}}),\label{eq:probability-w}
\end{equation}
 where $h_{{\rm eff}}=h(h/\Omega_{I})^{\alpha/(1-\alpha)}$ is a local
field fully renormalized by the baths according to (\ref{eq:new-h-adiabatic}).
If $w=1$, all interactions are larger than all fields, and the system
is in the conventional (strongly ordered) ferromagnetic phase. Conversely,
if $w=0$, all interactions are weaker than all transverse fields,
putting the system in the conventional (strongly disordered) paramagnetic
phase. Rare regions exist for $0<w<1$ with the inhomogeneously ordered
regime (the tail of the smeared transition) corresponding to $w\ll1$.

To leading approximation, the zero-temperature spontaneous magnetization
$m$ is simply the magnetic moment of the last spin cluster remaining
after the renormalization-group energy scale $\Omega$ has been iterated
to zero. In the tail of the smeared transition, $w\ll1$, this cluster
is made up of a collection of well separated rare regions of minimum
moment $1/\alpha$ on which the interactions are larger than the transverse
fields. The probability for finding a rare regions of $1/\alpha$
sites is simply $w^{1/\alpha}$. We thus conclude that the magnetization
in the tail of the smeared transition, $w\ll1$, behaves as 
\begin{equation}
m\sim w^{1/\alpha}.\label{eq:mag-onset}
\end{equation}
 Note that this result implies that the dependence of the magnetization
on the typical transverse field strength $h_{{\rm typ}}$ is nonuniversal;
it depends on the details of the bare distributions $P_{I}$ and $R_{I}$.
This nonuniversality agrees with heuristic results on other smeared
phase transitions.\ \cite{vojta-prl03,vojta-jpa-03,HNV11,SNHV12}

For weak dissipation and close to the location of the undamped quantum
critical point, the magnetization can also be estimated by the fraction
of undecimated spins when the undamped renormalization-group flow
reaches the crossover energy scale $T^{**}$ (because below $T^{**}$
almost no sites will be decimated). This can be readily accomplished
using the results of Ref.\ \onlinecite{fisher95}. The fraction of
undecimated spins at scale $T^{**}$ is given roughly by the product
of the density of surviving clusters at this scale $n(T^{**})$ and
their moment $\overline{\mu}(T^{**})$. As the moment is simply $1/\alpha$
(from the definition of $T^{**}$), we obtain 
\begin{equation}
m\sim\alpha^{1/(\phi\psi)-1}\label{eq:mag-xover-hc}
\end{equation}
 at the undamped quantum critical point, and 
\begin{equation}
m\sim\alpha^{-1}r^{\nu}\exp(-{\rm const}\times r^{1+\nu\psi\left(\phi-1\right)}/\alpha)\label{eq:mag-xover}
\end{equation}
 on its paramagnetic side. On the ferromagnetic side, the magnetization
is nonzero even in the absence of damping; weak dissipation thus provides
a small correction only. The result (\ref{eq:mag-xover}) holds sufficiently
close to the undamped quantum critical point. As the paramagnetic
phase is approached for larger $r$, the behavior crosses over to
(\ref{eq:mag-onset}) (see also Fig.\ 1 of Ref.\ \onlinecite{hoyos-vojta-prl08}).

The (order-parameter) magnetic susceptibility $\chi$ can be computed
by performing the renormalization group until the energy scale $\Omega$
reaches the temperature $T$. All spins decimated in this process
are rapidly fluctuating between up and down (i.e., pointing in the
$x$-direction due to their strong transverse field) and contribute
little to the susceptibility. All clusters remaining at scale $T$
are treated as free and contribute a Curie term $\mu^{2}/T$ to the
susceptibility.

For temperatures above $T^{**}$, the renormalization group flow coincides
with that of the undamped case. The temperature dependence of the
susceptibility is therefore identical to that of the dissipationless
transverse-field Ising chain.\ \cite{fisher95} In particular, in
the quantum critical region, $\chi$ increases as 
\begin{equation}
\chi\sim n(T)\overline{\mu}^{2}(T)/T\sim[\ln(\Omega_{I}/T)]^{2\phi-1/\psi}/T,
\end{equation}
 with vanishing $T$. In the paramagnetic Griffiths region, it behaves
as 
\begin{equation}
\chi\sim r^{\nu+2\nu\psi\left(1-\phi\right)}[\ln(\Omega_{I}/T)]^{2}\, T^{-1+1/z},
\end{equation}
 with $z\approx1/(2r)$ being the dynamical exponent.

Below $T^{**}$, the renormalization group flow deviates from the
undamped one, and the system crosses over into the inhomogeneously
ordered (IO) ferromagnetic phase. Here, the susceptibility can be
calculated from the fixed-point solution of Sec.\ \ref{subsub:ohmic}.
Following the same steps as in the \emph{ferromagnetic} Griffiths
phase of the undamped system, we obtain 
\begin{equation}
\chi\sim T^{-1-1/z^{\prime}},
\end{equation}
 with $z^{\prime}=1/{\cal P}_{0}$ being the low-energy dynamical
exponent in the ferromagnetic phase {[}see Eq.\ (\ref{eq:P-FP-ohmic}){]}.

It is desirable to relate the dynamical exponent $z^{\prime}$ of
the inhomogeneous ferromagnet to the bare distributions of $J$ and
$h$, or to the distance $r$ from the undamped quantum critical point.
The qualitative behavior is easily discussed. On the ferromagnetic
side of the undamped transition ($r<0$), $z^{\prime}$ approximately
agrees with its undamped value because dissipation is a subleading
correction, hence, $z^{\prime}\approx z\approx-1/(2r)$. In the inhomogeneously
ordered phase (i.e., on the paramagnetic side of the undamped transition,
$r>0$), ${\cal P}_{0}$ is determined by the distribution of the
interactions between the surviving clusters at energies below $T^{**}$.
With increasing transverse fields (increasing $r$), these clusters
become rarer implying that the distribution of their interactions
becomes broader. Thus, ${\cal P}_{0}$ decreases with increasing transverse
fields in the tail of the smeared transition. In the far tail, $w\ll1$,
the clusters surviving at energy scale $T^{**}$ are essentially independent
which means that their distances follow a Poisson distribution of
width $w^{-1/\alpha}$. This translates into the exponential distribution
(\ref{eq:P-FP-ohmic}) of the coupling variables $\eta$, with ${\cal P}_{0}\sim w^{1/\alpha}$.
Thus, the dynamical exponent 
\begin{equation}
z^{\prime}=1/{\cal P}_{0}\sim w^{-1/\alpha}\label{eq:z-prime}
\end{equation}
 diverges at the endpoint of the tail where $w=0$.

Close to the undamped quantum critical point, the behavior of $z^{\prime}$
can also be determined from the undamped renormalization group flow.
We approximate the flow in the Griffiths paramagnet by disregarding
the effects of dissipation until we reach the energy scale of $T^{**}$.
Below this scale, we assume that no further field decimations are
performed. The value of the parameter ${\cal P}_{0}$ can then be
obtained from the distribution ${\cal P}$ of the undamped Griffiths
paramagnet at the energy scale $\Gamma^{**}=\ln(\Omega_{I}/T^{**})$.
This yields 
\begin{equation}
z^{\prime}\sim\frac{1}{r}\exp[{\rm const}\times r^{\psi\nu\left(\phi-1\right)+1}/\alpha].\label{eq:z-FM}
\end{equation}
 Thus, $z^{\prime}$ increases with increasing $r$, in agreement
with the above heuristic arguments. Our results for $z^{\prime}$
imply a dramatic change of the dynamical exponent when crossing over
from the quantum Griffiths paramagnet to the inhomogeneously ordered
ferromagnet at $T^{**}$. Such behavior was indeed observed by means
of a numerical implementation of the renormalization-group rules.\ \cite{vojta-hoyos-procee-dissip,hoyos-unpublished}

\section{Discussion \label{sec:Discussions}}

\subsection{Applicability to weakly disordered systems}

In this section, we discuss to what extent our results apply to systems
with weak or moderate bare disorder. The strong-disorder renormalization-group
recursion relations derived in Sec.\ \ref{sec:RG-recursion-relation}
become asymptotically exact in the limit of infinite disorder, but
they are only approximations for finite disorder. For this reason,
the strong-disorder renormalization group needs to be complemented
by other methods to discern the fate of weakly disordered systems.

To analyze the limit of weak disorder, we can start from the Harris
criterion,\ \cite{harris-jpc74} which states that a clean critical
point is stable against weak disorder if its correlation length exponent
$\nu$ fulfills the inequality $d\nu>2$. The clean Ohmic transverse-field
Ising chain features a quantum critical point with an exponent value
$\nu\approx0.638$.\ \cite{werner-etal-prl-05} Disorder is thus
a relevant perturbation. Kirkpatrick and Belitz\ \cite{kirkpatrick-belitz-prl96}
and Narayanan \emph{et al}.\ \cite{narayanan-vojta-belitz-kirkpatrick-prl99,narayanan-vojta-belitz-kirkpatrick-prb99}
studied a quantum order-parameter field theory with Ohmic dissipation
by means of a conventional perturbative renormalization group. This
method, which is controlled at \emph{weak} disorder, did not produce
a stable critical fixed point, but runaway flow toward large disorder
(where the strong-disorder renormalization group becomes better and
better). This strongly suggests that our results apply for any amount
of (bare) disorder.

We now turn to the question of whether or not our results become asymptotically
exact in the low-energy limit. To address this question, we need to
discuss the widths of the distributions $\pi$ and $\rho$ of the
logarithmic interaction and transverse-field variables in the inhomogeneously
ordered ferromagnet. Using the fixed point distributions (\ref{eq:P-FP-ohmic})
and (\ref{eq:R-FP-ohmic}) and the corresponding scale factors $f_{\zeta}=1$
and $f_{\beta}=\Gamma e^{{\cal P}_{0}\Gamma}$, we find the widths
to be given by $1/\pi_{0}=1/{\cal P}_{0}$ and $1/\rho_{\beta,0}=\Gamma e^{{\cal P}_{0}\Gamma}/{\cal R}_{\theta,0}$.
Thus, the relative width of the transverse-field distribution diverges,
while the width of the interaction distribution remains finite in
the limit $\Gamma\to\infty$ ($\Omega\to0$).

This result can be easily understood by following the renormalization-group
flow. Above the crossover energy scale $\Omega^{**}$ (where the typical
cluster size reaches $1/\alpha$), the flow approximately coincides
with that of a paramagnetic Griffiths phase of the undamped problem.
Thus, (almost) only sites are decimated, which enormously broadens
the distribution $\pi$ of the interactions. Below $\Omega^{**}$,
almost all decimations are interaction decimations. Therefore, $\pi$
does not renormalize further, and its width remains large but finite.
In contrast, the transverse fields are driven to zero and their distribution
broadens without limit. Our results for the inhomogeneously ordered
ferromagnet are thus not asymptotically exact, but represent a good
approximation. However, in the tail of the smeared transition, $w\to0$,
the width of the interaction distribution diverges {[}see Eqs.\ (\ref{eq:z-prime})
and (\ref{eq:z-FM}){]}.

In this large-disorder limit, higher-order corrections to the recursion
relations derived in Sec.\ \ref{sec:RG-recursion-relation} are suppressed.
This includes couplings between different dissipative baths that appear
upon integrating out a site. As these couplings only appear in fourth
order of perturbation theory, they are very weak (and of short-range
type) in the large-disorder limit. Thus, all corrections to our recursion
relations are irrelevant in the renormalization-group sense, and our
theory becomes asymptotically exact in the tail of the smeared transition.

Finally, we discuss to what extent our results would change if the
bare system had long-range interactions between the different baths
(or equivalently, a single global bath). In this case, the true zero-temperature
behavior would be dominated by these interactions and differ from
our results. However, sufficiently weak long-range couplings would
become important only at very low temperatures {[}below even $T^{**}$,
see Fig.\ \hyperref[fig:PD]{\ref{fig:PD}(b)}{]}. Thus, our theory
would remain valid in a broad temperature window. The precise value
of the crossover energy scale depends on the bare value of the long-range
interactions and on in the details of the model. It is a nonuniversal
quantity and beyond the scope of this work.

\subsection{Sub-Ohmic dissipation}

As discussed in Sec.\ \ref{subsub:sub-Ohmic-flow}, our strong-disorder
renormalization-group method can not be directly applied to the sub-Ohmic
case because there is no phase transition if the dissipative baths
are treated within adiabatic renormalization. However, a qualitative
picture of the physics of the sub-Ohmic case can nonetheless be developed.

The sub-Ohmic spin-boson model is known\ \cite{kehrein-mielke-pla96,bulla-tong-vojta-prl03}
to undergo a continuous quantum phase transition from a fluctuating
phase to a localized (frozen) phase as the transverse field is decreased
(or the dissipation strength increased). Thus, the dynamics of sufficiently
large spin clusters (which have small effective transverse fields
and large dissipation strength) is always frozen. These frozen (classical)
clusters can be aligned by any infinitesimal interaction, leading
to magnetic long-range order. Just as in the Ohmic case, this mechanism
replaces the paramagnetic Griffiths phase and the quantum critical
point by an inhomogeneously ordered ferromagnetic phase and smears
the underlying ferromagnetic quantum phase transition.

\subsection{Higher dimensions}

We now discuss to what extent our results also apply to dissipative
random transverse-field Ising models in higher space dimensions. The
renormalization-group steps of Sec.\ \ref{sec:RG-recursion-relation}
can be performed in complete analogy to the one-dimensional case,
and the resulting recursion relations take the same form as Eqs.\ (\ref{eq:new-h-adiabatic}),
(\ref{eq:J-tilde}), (\ref{eq:h-tilde}), and (\ref{eq:mu-tilde}).
However, decimating a site now generates effective interactions between
all pairs of its neighboring sites, while decimating an interaction
leads to an effective site (cluster) that couples to a larger number
of neighbors than before. Thus, both decimation steps change the connectivity
of the lattice. Moreover, they introduce correlations between the
remaining couplings. For this reason, an analytical treatment of the
renormalization group flow appears to be impossible, even in the dissipationless
case. However, the strong-disorder renormalization group has been
implemented numerically in two and higher dimensions by keeping track
of the irregular ``maze'' of new interactions created by the decimation
steps.\ \cite{motrunich-ising2d,kovacs-igloi-prb11} An analogous
approach would be possible in the dissipative case.

Even in the absence of a complete solution, the effects of dissipation
on higher-dimensional random transverse-field Ising models can be
understood qualitatively. The extra downward renormalization (\ref{eq:new-h-adiabatic})
of the transverse fields due to the baths is purely local, and thus
also applies to higher dimensions. Just as in one dimension, it reduces
the probability of decimating a site, $(1-\alpha\overline{\mu}_{\Omega})R_{h}(\Omega)$.
Once the moment $\mu$ of a cluster reaches $1/\alpha$, it will never
be decimated. We conclude that clusters with moments $\mu>1/\alpha$
will freeze implying that a critical fixed-point solution is impossible.
As in one dimension, the quantum phase transition is therefore smeared,
and the paramagnetic Griffiths phase is replaced by an inhomogeneously
ordered ferromagnetic phase.

Although the zero-temperature quantum phase transition of the higher-dimensional
random transverse-field Ising models is similar to the one-dimensional
case, their behavior at nonzero temperatures differs. In one dimension,
long-range order only exists at zero temperature (see our phase diagrams,
Fig.\ \ref{fig:PD}). In higher dimensions, a long-range ordered
ferromagnetic phase can exist at nonzero temperatures. It is important
to note that the classical phase transition separating this phase
from the paramagnet at nonzero temperatures is \emph{not} smeared.
At nonzero temperatures, even the largest clusters (rare regions)
will thermally fluctuate. Thus, a finite interaction is required to
align them. This restores a sharp phase transition.\ \cite{vojta-prl03}

In two or more space dimensions, disorder can also be introduced into
the (dissipationless) transverse-field Ising model simply by diluting
the lattice. For weak transverse fields, the ferromagnetic phase survives
up to the percolation threshold $p_{c}$ of the lattice. The quantum
percolation phase transition emerging at $p_{c}$ shares many characteristics
with the infinite-randomness critical points of the generic random
transverse-field Ising models.\ \cite{senthil-sachdev-prl96} Adding
Ohmic dissipation suppresses the quantum fluctuations of sufficiently
large percolation clusters. The results is an unusual classical super-paramagnetic
phase.\ \cite{hoyos-vojta-prb06} However, the percolation transition
remains sharp as it is driven by the critical lattice geometry at
the percolation threshold.

\subsection{Experiments }

To the best of our knowledge, an experimental realization of the \emph{microscopic}
Hamiltonian defined in Eqs.\ (\ref{eq:H}) to (\ref{eq:H-coupling})
has not been found yet. However, its order-parameter field theory,
a one-component $\phi^{4}$ theory with an inverse Gaussian propagator
of the form $G(\mathbf{q},\omega)^{-1}=r+\mathbf{q}^{2}+|\omega|^{s}$,
also describes a number of experimentally important quantum phase
transitions. Based on universality, we expect the qualitative properties
of these transitions to be analogous to those of our model (\ref{eq:H}).

For example, Hertz' theory\ \cite{hertz-prb76} of the antiferromagnetic
quantum phase transition of itinerant electrons leads to such an order-parameter
field theory. In this case, the dissipation is Ohmic ($s=1$); it
is caused by the electronic particle-hole excitations rather than
abstract heat baths. For disordered itinerant ferromagnets, the dynamic
part of the Gaussian propagator takes the form $|\omega|/\mathbf{q}^{2}$
rather than $|\omega|$. Experiments on these disordered metallic
quantum magnets often show unusual behavoir,\ \cite{stewart-rmp01,stewart-addendum-rmp06,lohneysen-etal-rmp07}
and strong-disorder effects have been suggested as possible reasons.\ \cite{castroneto-jones-prb00}
However, explicit verifications of the quantum Griffiths and smeared-transition
scenarios were missing for a long time. Only recently, some promising
results have been found in several itinerant ferromagnets.\ \cite{guo-etal-prl08,westercamp-etal-prl09,ubaidkassis-vojta-schroeder-prl10}
For a more thorough discussion, the reader is referred to Refs.\ \onlinecite{vojta-review06,vojta-jltp-10}.

We emphasize, that even though the leading terms in the order-parameter
field theory of the dissipative random transverse-field Ising model
(\ref{eq:H}) agree with the corresponding terms in itinerant quantum
magnets, their behaviors are not completely identical. Importantly,
in the Hamiltonian (\ref{eq:H}), each spin couples to its own independent
dissipative bath. In contrast, in metallic magnets, all spins couple
to the same Fermi sea. This produces an additional long-range interaction
between the spins which has no counterpart in the Hamiltonian (\ref{eq:H}).
It is mediated by the fermionic particle-hole excitations and known
as the Ruderman-Kittel-Kasuya-Yosida (RKKY) interaction.

\section{Conclusions \label{sec:Conclusions}}

To summarize, we have studied the quantum phase transition of the
dissipative random transverse-field Ising chain via an analytical
implementation of a strong-disorder renormalization-group method.
We have shown that super-Ohmic dissipation is an irrelevant perturbation.
The quantum phase transition is thus in the same universality class
as the dissipationless chain, i.e., it is governed by an infinite-randomness
fixed point and accompanied by quantum Griffiths singularities.

In contrast, for Ohmic dissipation, the sharp quantum phase transition
is destroyed by smearing because sufficiently large rare regions completely
freeze at zero temperature. Note that this as a consequence of the
\emph{interplay} between disorder and dissipation as neither disorder
nor dissipation alone can smear the quantum phase transition of the
transverse-field Ising chain. The behavior of the sub-Ohmic case is
qualitatively similar to the Ohmic one: it also leads to a smeared
transition.

The results of this paper apply to the case of discrete Ising order-parameter
symmetry. Systems with continuous $O(N)$ order-parameter symmetry
behave differently. In contrast to sufficiently large Ising clusters,
which freeze in the presence of Ohmic dissipation, $O(N)$ clusters
of all sizes continue to fluctuate with a rate that depends exponentially
on their magnetic moment.\ \cite{vojta-schmalian-prb05} The resulting
sharp phase transition is controlled by an infinite-randomness critical
point in the universality class of the undamped random transverse-field
Ising model.\ \cite{hoyos-kotabage-vojta-prl07,vojta-kotabage-hoyos-prb09}
Super-Ohmic dissipation has weaker effects,\ \cite{vojta-etal-jp011}
and the renormalization-group flow is very similar to the dissipationless
case.\ \cite{altman-etal-prl04}

The difference between the discrete and continuous symmetry cases
as well as the differences between the different types of dissipation
can all be understood in terms of a classification of rare-region
effects according to the effective defect (rare-region) dimensionality\ \cite{vojta-schmalian-prb05,vojta-review06}:
If the rare-region dimensionality $d_{RR}$ is below the lower critical
dimension $d_{c}^{-}$ of the problem at hand, the transition is sharp
and conventional. If $d_{RR}=d_{c}^{-}$, the transition is still
sharp but controlled by an infinite-randomness critical point. Finally,
if finite-size rare regions can order independently of each other
($d_{RR}\ge d_{c}^{-}$), the transition is smeared. This classification
applies to classical phase transitions and to quantum phase transitions
that can be related to classical ones via the quantum-to-classical
mapping.\ \cite{sachdev-book}

We expect our work or appropriate generalizations to shed light onto
a variety of quantum phase transitions in disordered systems in which
the order-parameter modes are coupled to additional noncritical soft
modes.
\begin{acknowledgments}
This work has been supported in part by the NSF under grant No. DMR-0906566,
Research Corporation, FAPESP under Grant No. 2010/ 03749-4, and CNPq
under grants No. 590093/2011-8 and No. 302301/2009-7.
\end{acknowledgments}
\appendix

\section{The RG flow in the Ohmic ferromagnetic phase\label{sec:Flow-equation-FM}}

In this appendix, the flow equations (\ref{eq:flow-rescale-P}) and
(\ref{eq:flow-rescale-R}) are studied in the ferromagnetic regime
for Ohmic ($s=1$) damping.

For ferromagnetic fixed-point solutions, the convolution term in (\ref{eq:flow-rescale-P})
has to drop out, leading to 
\begin{equation}
0=\frac{\dot{f}_{\zeta}}{f_{\zeta}}\left({\cal P}+\eta\frac{\partial{\cal P}}{\partial\eta}\right)+\frac{1}{f_{\zeta}}\left(\frac{\partial{\cal P}}{\partial\eta}+{\cal P}_{0}{\cal P}\right).\label{eq:flow-P-reduced}
\end{equation}
 As no sites are decimated, the typical value of $\zeta$ cannot grow
(as in the undamped case). Thus, $f_{\zeta}=1$ and 
\begin{equation}
{\cal P}^{*}(\eta)={\cal P}_{0}e^{-{\cal P}_{0}\eta},
\end{equation}
 with ${\cal P}_{0}$ being an integration constant. With $f_{\zeta}=1$
and assuming that 
\begin{equation}
f_{\mu}/f_{\beta}\rightarrow0\label{eq:fmu-fbeta}
\end{equation}
 (because we expect that $f_{\beta}$ grows faster than $f_{\mu}$
due to the dissipation), the fixed-point equation (\ref{eq:flow-rescale-R})
reads 
\begin{align}
0= & \frac{\dot{f}_{\beta}}{f_{\beta}}\left({\cal R}+\theta\frac{\partial{\cal R}}{\partial\theta}\right)+\frac{\dot{f}_{\mu}}{f_{\mu}}\left({\cal R}+\nu\frac{\partial{\cal R}}{\partial\nu}\right)\nonumber \\
 & +{\cal P}_{0}\left({\cal R}\stackrel{\theta,\nu}{\otimes}{\cal R}-{\cal R}\right).\label{eq:flow-R-reduced-1}
\end{align}
 Interestingly, all dissipative terms (those involving $\alpha$)
drop out, and we recover the undamped fixed-point equation. Therefore,
we can integrate either over the fields or over the magnetic moments
and analyze ${\cal R}_{\theta}$ or ${\cal R}_{\nu}$ separately.
Physically relevant solutions require that all terms in (\ref{eq:flow-R-reduced-1})
survive in the limit $\Gamma\to\infty$ which implies $\dot{f}_{\beta}/f_{\beta}=\dot{f}_{\mu}/f_{\mu}={\rm const}$.
Integrating over $\nu$ and Laplace transforming (\ref{eq:flow-R-reduced-1}),
we arrive at 
\begin{equation}
0=-cz\frac{\partial}{\partial z}\hat{{\cal R}}^{*}(z,0)+{\cal P}_{0}\hat{{\cal R}}^{*}(z,0)\left(\hat{{\cal R}}^{*}(z,0)-1\right),\label{eq:LT-R-nu}
\end{equation}
 where $\hat{{\cal R}}(v,n)=\int_{0}^{\infty}e^{-v\theta-n\nu}{\cal R}(\theta,\nu){\rm d}\theta{\rm d}\nu$
and $c\geq{\cal P}_{0}$ is a constant. Equation (\ref{eq:LT-R-nu}),
which can be linearized via $\hat{{\cal R}}^{*}=-c(\partial_{\ln v}\ln u)/{\cal P}_{0}$,
then yields 
\begin{equation}
\hat{{\cal R}}^{*}(v,0)=\frac{1}{1+\left(v/A\right)^{{\cal P}_{0}/c}},\label{eq:FP-LT-R}
\end{equation}
 where $A$ is an integration constant. For $c\neq{\cal P}_{0}$,
the inverse Laplace transformation gives ${\cal R}_{\theta}^{*}(\theta)\propto\theta^{-1-{\cal P}_{0}/c}$
for $\theta\gg1$, which corresponds to $R_{h}^{*}(h)\sim1/(h|\ln h|^{1+{\cal P}_{0}/c})$
for $h\ll1$. Although this extremely singular distribution is a fixed-point
solution of the flow equations, it is \emph{not} the attractive one
for typical initial distributions. Indeed, this extreme fixed point
can only be accessed if the \emph{bare} distribution of fields already
contains such strong singularity.\ \cite{fisher95}

The less singular (and attractive) solutions correspond to $c={\cal P}_{0}$.
By inverse Laplace transforming Eq.\ (\ref{eq:FP-LT-R}), we conclude
that ${\cal R}_{\theta}^{*}={\cal R}_{\theta0}e^{-{\cal R}_{\theta0}\theta}$,
which is identical to the undamped case.\ \cite{fisher95} The same
analysis applies to ${\cal R}_{\nu}^{*}={\cal R}_{\nu0}e^{-{\cal R}_{\nu0}\nu}$.
Here, ${\cal R}_{\theta0}$ and ${\cal R}_{\nu0}$ are nonuniversal
constants. Moreover, since the flow of the magnetic moments is identical
to the undamped case, we conclude that $f_{\mu}=e^{{\cal P}_{0}\Gamma}$.
Recall we cannot use the result of the undamped case in which $f_{\beta}=f_{\mu}$
because of our assumption (\ref{eq:fmu-fbeta}).

After finding the reduced fixed-point distributions ${\cal R}_{\theta}^{*}$
and ${\cal R}_{\nu}^{*}$, we now search for the joint fixed-point
distribution ${\cal R}^{*}(\theta,\nu)$ with $\dot{f}_{\beta}/f_{\beta}=\dot{f}_{\mu}/f_{\mu}={\cal P}_{0}$
and $f_{\mu}/f_{\beta}=0$ in the limit $\Gamma\rightarrow\infty$.
The Laplace-transformed flow equation (\ref{eq:flow-R-reduced-1})
becomes 
\begin{equation}
0=-v\frac{\partial\hat{{\cal R}}}{\partial v}-n\frac{\partial\hat{{\cal R}}}{\partial n}+\hat{{\cal R}}\left(\hat{{\cal R}}-1\right),\label{eq:dif-eq-R-hat}
\end{equation}
 which can be linearized via $\hat{{\cal R}}=-\partial_{m}\ln u$,
where $m=\ln v+\ln n$. Hence, 
\begin{equation}
\hat{{\cal R}}^{*}(v,n)=\left(1+f(v/n)\sqrt{vn}\right)^{-1},\label{eq:FP-R-1}
\end{equation}
 where $f(k)$ is a generic function constrained to diverge $\sim\sqrt{k}/{\cal R}_{\theta0}$
when $k\rightarrow\infty$, and to diverge $\sim1/(\sqrt{k}{\cal R}_{\nu0})$
when $k\rightarrow0$ as dictated by the solution (\ref{eq:FP-LT-R}).

Our purpose now is to find the correct function $f(k)$ in (\ref{eq:FP-LT-R}).
It is easy to see that there are two limiting cases of the joint distribution
${\cal R}(v,n)$ that are compatible with the reduced distributions
${\cal R}_{\theta}^{*}$ and ${\cal R}_{\nu}^{*}$ found above, viz.,
\begin{equation}
\hat{{\cal R}}_{{\rm corr}}(v,n)=\left(1+\frac{v}{R_{\theta0}}+\frac{n}{R_{\nu0}}\right)^{-1},\label{eq:correlated}
\end{equation}
 and 
\begin{equation}
\hat{{\cal R}}_{{\rm uncorr}}(v,n)=\left(1+\frac{v}{R_{\theta0}}\right)^{-1}\left(1+\frac{n}{R_{\nu0}}\right)^{-1},\label{eq:uncorrelated}
\end{equation}
 which, respectively, correspond to perfect correlations, ${\cal R}_{{\rm corr}}(\theta,\nu)=R_{\theta0}R_{\nu0}e^{-R_{\theta0}\theta}\delta(R_{\theta0}\theta-R_{\nu0}\nu)$,
and no correlations, ${\cal R}_{un{\rm corr}}(\theta,\nu)=R_{\theta0}e^{-R_{\theta0}\theta}R_{\nu0}e^{-R_{\nu0}\nu}$,
between transverse fields and magnetic moments. However, $\hat{{\cal R}}_{{\rm uncorr}}$
does \emph{not} satisfy the general form (\ref{eq:FP-R-1}). This
suggests that correlations between the fields and magnetic moments
arise from the fixed-point structure itself, regardless the strength
of the damping $\alpha$. Indeed, as we shall see below, the flow
is toward the perfect correlated case (\ref{eq:correlated}).

Since it is not possible to gather full information about $f(v/n)$
by analyzing the fixed-point solution alone, we now analyze the renormalization-group
flow near the fixed point. This can only be accomplished if the $\Gamma$-dependence
is not scaled out. We thus consider the unscaled variables $\beta$
and $\mu$ and use the fixed point (\ref{eq:FP-R-1}) as reference.
Having in mind that $f_{\mu}=e^{{\cal P}_{0}\Gamma}$, $f_{\mu}/f_{\beta}\to0$,
and $\dot{f}_{\beta}/f_{\beta}\rightarrow{\cal P}_{0}$, we keep only
the main terms as well as the leading corrections in the flow equation
(\ref{eq:flow-rho}): 
\begin{equation}
\frac{\partial\rho}{\partial\Gamma}=-\alpha\mu\frac{\partial\rho}{\partial\beta}+\pi_{0}\left(\rho\otimes\rho-\rho\right).\label{eq:flow-rho-FM}
\end{equation}
 It is easy to see that the neglected term corresponds to the third
one on the right-hand side of (\ref{eq:flow-rescale-R}), which is
the one responsible for the normalization of ${\cal R}$ due to a
field decimation. The effects of damping thus enter only in the renormalization
of the fields upon cooling the bath.

The careful reader may notice that the neglected term is proportional
to $f_{\mu}/f_{\beta}$ as is the fourth one. Then, why is $-\alpha\overline{\mu}_{0}\rho_{\beta,0}\rho$
neglected and $-\alpha\mu\partial\rho/\partial\beta$ is not? The
reason comes from the arising correlations between transverse fields
and magnetic moments. Consider for instance the perfectly correlated
case ${\cal R}_{{\rm corr}}$. With this choice, 
\[
\overline{\mu}_{0}=f_{\mu}\overline{\nu}_{0}=f_{\mu}\frac{\int\nu{\cal R}_{{\rm corr}}(0,\nu){\rm d}\nu}{\int{\cal R}_{{\rm corr}}(0,\nu){\rm d}\nu}=0.
\]
 This result is interesting because, as argued in Sec.\ \ref{sec:Flow-equations},
$(1-\alpha\overline{\mu}_{0}\rho_{\beta,0})\rho$ is the probability
of having a decimation of a field. Since the magnetic moments grow
$\sim f_{\mu}$ along the renormalization group flow, one might naively
expect that the mean magnetic moment to be decimated, $\overline{\mu}_{0}$,
also grows $\sim f_{\mu}$. This would imply that the above probability
becomes negative (notice $\rho_{\beta,0}$ is nonzero). Therefore,
correlations between the fields and moments ensure that this probability
remains well-defined.

By Laplace-transforming (\ref{eq:flow-rho-FM}), we finally arrive
at 
\begin{equation}
\frac{\partial\hat{\rho}}{\partial\Gamma}=\alpha b\frac{\partial\hat{\rho}}{\partial m}+\pi_{0}\left(\hat{\rho}^{2}-\hat{\rho}\right),\label{eq:LT-flow-rho-FM}
\end{equation}
 where $\hat{\rho}(b,m)=\int e^{-b\beta}\rho^{\prime}(\beta,m){\rm d}\beta$
and $\rho^{\prime}(\beta,m)=\int e^{-m\mu}\rho(\beta,\mu){\rm d}\mu$.
(The term $\alpha\partial\rho^{\prime}(0,m)/\partial m$ also vanishes
because of the correlations between fields and magnetic moments.)
The first term on the right-hand side of (\ref{eq:LT-flow-rho-FM})
is the desired leading correction to the fixed-point flow equation
(\ref{eq:dif-eq-R-hat}). Inspection of (\ref{eq:LT-flow-rho-FM}),
guided by the fixed-point result (\ref{eq:FP-R-1}), yields 
\[
\hat{\rho}^{*}(b,m;\Gamma)=\left[1+F\left(\Gamma+\frac{m}{\alpha b},b\right)\exp\left(-\pi_{0}\frac{m}{\alpha b}\right)\right]^{-1},
\]
 where $F(y,b)$ is an integration constant which depends on $b$
and on the combination $y=\Gamma+\frac{m}{\alpha b}$, which implies
$m/b\propto\Gamma$, i.e., $f_{\beta}/f_{\mu}\propto\Gamma$. To see
this clearly, let us compute $F$ from some initial conditions.

Consider firstly a perfectly correlated one, i.e., 
\begin{equation}
\hat{\rho}(b,m;0)\equiv\hat{\rho}_{{\rm corr}}=\left(1+\frac{b}{\rho_{\beta0}}+\frac{m}{\rho_{\mu0}}\right)^{-1}.\label{eq:IC-correlated}
\end{equation}
 In this case $F(y,b)=b(1+\alpha\rho_{\beta0}y/\rho_{\mu0})e^{\pi_{0}y}/\rho_{\beta0}$.
Thus, $\hat{\rho}^{*}(b,m;\Gamma)=\hat{\rho}(b,m;\Gamma\rightarrow\infty)$
with 
\begin{equation}
\hat{\rho}^{*}(b,m;\Gamma)=\left(1+\frac{\alpha f_{\beta}b+f_{\mu}m}{\rho_{\mu0}}\right)^{-1},\label{eq:FP-LT-rho-FM}
\end{equation}
 with $f_{\mu}=e^{\pi_{0}\Gamma}$ as expected, and $f_{\beta}=\Gamma e^{\pi_{0}\Gamma}(1+{\cal O}(\alpha\Gamma)^{-1})$.
Notice that $\hat{\rho}^{*}$ is of the perfectly correlated type
(\ref{eq:correlated}). Finally, we obtain the scale factor $f_{\beta}\sim\Gamma f_{\mu}$
which satisfies the requirements that $f_{\mu}/f_{\beta}\rightarrow0$
and $\dot{f}_{\mu}/f_{\mu}=\dot{f}_{\beta}/f_{\beta}={\cal P}_{0}=\pi_{0}$
in the limit $\Gamma\rightarrow\infty$.

To convince ourselves that $\hat{\rho}^{*}$ is the relevant attractive
fixed-point distribution, consider, for example, an initial condition
which is totally at odds with (\ref{eq:IC-correlated}): the perfectly
uncorrelated case 
\[
\hat{\rho}(b,m;0)\equiv\hat{\rho}_{{\rm uncorr}}=\left(1+\frac{b}{\rho_{\beta0}}\right)^{-1}\left(1+\frac{m}{\rho_{\mu0}}\right)^{-1}.
\]
 In this case, $F(y,b)=b[1+\alpha y(\rho_{\beta0}+b)/\rho_{\mu0}]e^{\pi_{0}y}/\rho_{\beta0}$
which yields 
\[
\hat{\rho}(b,m;\Gamma)=\left(1+\frac{\alpha f_{\beta}b+f_{\mu}m}{\rho_{\mu0}}+{\cal O}\left(b,\frac{1}{\Gamma}\right)\right)^{-1}.
\]
 In the limit $\Gamma\rightarrow\infty$, and $b,m\rightarrow0$,
this distribution flows to the attractive solution $\hat{\rho}^{*}(b,m;\Gamma)$
(\ref{eq:FP-LT-rho-FM}). Since both uncorrelated and perfectly correlated
initial conditions flow to the same (correlated) fixed-point solution
(\ref{eq:FP-LT-rho-FM}), we expect all well-behaved (ferromagnetic)
initial distributions to flow toward this fixed point. We have also
checked this numerically in Ref.\ \onlinecite{vojta-hoyos-procee-dissip}.
After the inverse Laplace transformation, we finally obtain 
\begin{equation}
\rho^{*}(\beta,\mu;\Gamma)=\rho_{\beta0}\rho_{\mu0}e^{-\rho_{\beta0}\beta}\delta(\rho_{\beta0}\beta-\rho_{\mu0}\mu),\label{eq:FR-rho-FM}
\end{equation}
 with $\rho_{\mu0}=\rho_{0}/f_{\mu}$ and $\rho_{\beta0}=\rho_{0}/(\alpha f_{\beta})$.

\bibliographystyle{apsrev4-1}
\bibliography{referencias}

\begin{thebibliography}{10}
\bibitem{footnote1}%
  \BibitemOpen
  \bibinfo {note} {If $\dot{f}_{\zeta}/f_{\zeta}$ vanishes more slowly than
  $1/f_{\zeta}$, then the first term on the right-hand side of 
  Eq.~(\ref{eq:flow-rescale-P}) is the leading one yielding ${\cal
  P}^{*}=\eta^{-1}$ which is not normalizable. If, in turn, $1/f_{\zeta}$
  vanishes more slowly, then the second and third terms on the right-hand side
  of Eq.~(\ref{eq:flow-rescale-P}) become the leading ones. After a Laplace
  transformation, one finds that ${\cal P}^{*}=J_{1}(2P_{0}\eta)/(2\eta)$, with
  $J_{1}$ being the Bessel function of first kind, which has nonphysical
  negative values.}%
  \bibAnnoteFile{Stop}{footnote1}%
\end{thebibliography}

\end{document}